\documentclass[journal]{IEEEtran}
\ifCLASSINFOpdf
\else
\fi
\usepackage{booktabs}
\usepackage{graphicx} 
\usepackage{stfloats}

\usepackage{textcomp}
\usepackage{stfloats}
\usepackage{verbatim}
\usepackage{graphicx}
\usepackage{booktabs}
\usepackage{amsmath,amsfonts}
\usepackage{algorithmic}
\usepackage{array}
\usepackage{stfloats}
\usepackage{verbatim}
\usepackage{graphicx}
\usepackage{balance}
\usepackage{balance}
\usepackage[utf8]{inputenc}
\usepackage{caption}
\usepackage{graphicx}
\usepackage{float} 
\usepackage{subcaption}
\usepackage{color}
\usepackage{makecell}

\newcommand{\specialcell}[2][c]{%
	\begin{tabular}[#1]{@{}l@{}}#2\end{tabular}}

\begin{document}

\title{A Fine-Grained 3D Radio Map Construction Paradigm with Ultra-Low Sampling Rates by\\Large Generative Models}
\author{Zhiyuan~Liu,~\IEEEmembership{Student Member,~IEEE, }
Qingyu~Liu,~\IEEEmembership{Member,~IEEE, }
Shuhang~Zhang,~\IEEEmembership{Member,~IEEE, }
Hongliang~Zhang,~\IEEEmembership{Member,~IEEE, }  Lingyang~Song,~\IEEEmembership{Fellow,~IEEE}
\thanks{Received 14 May 2025; revised 13 October 2025 and 11 January 2026; accepted 21 March 2026.
This work was supported in part by the National Natural Science Foundation of China under Grant 62527801, 62401025, and 62401302; in part by the GuangDong Basic and Applied Basic Research Foundation under Grant 2023A1515110120; in part by the Key Project of Beijing Natural Science Foundation under Grant L253004.
\emph{(Corresponding authors: Qingyu Liu; Lingyang Song)}
}
\thanks{Zhiyuan Liu is with the School of Electronic and Computer Engineering,  Peking University Shenzhen Graduate School, Shenzhen, China, 518055 (e-mail:
{liuzhiyuan@stu.pku.edu.cn}).}
\thanks{Qingyu Liu is with the School of Electronic and Computer Engineering, Peking University Shenzhen Graduate School, Shenzhen, China, 518055, and also with the Peng Cheng Laboratory, Shenzhen, China, 518055 (e-mail: {qy.liu@pku.edu.cn}).}
\thanks{Shuhang Zhang and Hongliang Zhang are with the State Key Laboratory of Photonics and Communications, School of Electronics, Peking University, Beijing, China, 100871 (email: 
{zhangshuhang@pku.edu.cn}, {hongliang.zhang@pku.edu.cn}).}
\thanks{Lingyang Song is with the State Key Laboratory of Photonics and Communications, School of Electronics, Peking University, Beijing, China, 100871, also with the School of Electronic and Computer Engineering, Peking University Shenzhen Graduate School, Shenzhen, China, 518055, and also with the Peng Cheng Laboratory, Shenzhen, China, 518055 (e-mail: 
{lingyang.song@pku.edu.cn}).}
\thanks{{Code is publicly available at: https://github.com/liuzhiyuan-pku/RadioLAM}}}

\maketitle
\thispagestyle{empty}
\pagestyle{empty}

\begin{abstract}
A radio map captures the spatial distribution of wireless channel parameters, such as the strength of the signal received, across a geographic area.
The problem of fine-grained three-dimensional (3D) radio map construction involves inferring a high-resolution radio map for the two-dimensional (2D) area at an arbitrary target height within a 3D region of interest, using radio samples collected by sensors sparsely distributed in that 3D region.
Solutions to the problem are crucial for efficient spectrum management in 3D spaces, particularly for drones in the rapidly developing low-altitude economy.
However, this problem is challenging due to ultra-sparse sampling, where the number of collected radio samples is far fewer than the desired resolution of the radio map to be estimated.
In this paper, we design {RadioLAM, a fine-grained 3D radio map construction paradigm built on generative Large Artificial Intelligence Models (LAMs)}.
RadioLAM employs the creative power and the strong generalization capability of LAM to address the ultra-sparse sampling challenge.
It consists of three key blocks:
1) an augmentation block, using the radio propagation model to project the radio samples collected at different heights to the 2D area at the target height;
2) a generation block, leveraging a diffusion-based LAM under an Mixture of Experts (MoE) architecture to generate a candidate set of fine-grained radio maps for the target 2D area;
and 3) an election block, utilizing the radio propagation model as a guide to find the best map from the candidate set.
Extensive simulations show that RadioLAM is able to solve the fine-grained 3D radio map construction problem efficiently from an ultra-low sampling rate of $0.1\%$, and significantly outperforms state-of-the-art (SOTA).
{Furthermore, real-world experiments confirm that RadioLAM achieves superior performance compared to SOTA.}
\end{abstract}


\section{Introduction}
\subsection{Background}
A radio map presents parameters of interest in wireless communication channels, such as the received signal strength (RSS), at each point of a certain geographical area~\cite{RME}.
A fine-grained radio map is a high-resolution radio map for a large-scale geographical area.
It can enhance the performance of many wireless applications by enabling dynamic spectrum access \cite{Peng24:TCE:Access}, efficient spectrum sharing~\cite{Matar24:JSAC:Sharing}, and intelligent interference management~\cite{Huang24:TWC:inference}.
Of particular importance are the fine-grained radio map for a three-dimensional (3D) area, which can significantly improve the spectrum utilization of growing number of low-flying aircrafts in the 3D airspace, and is critical for supporting emerging low-altitude economy applications, e.g., aerial surveillance~\cite{Zhu24:IoTJ:surveillance}, search \& rescue (SAR)~\cite{Qi24:TVT:SAR}, and drone-assisted infrastructure inspection~\cite{Ri24:NC:inspection}.

Radio map construction problems can generally be categorized into two classes: the sampling-free problem and the sampling-based problem. 
The sampling-free problem uses the information from the transmitters (base stations) to estimate radio maps~\cite{ray-tracing,dominant-path,eg2,eg}.
It assumes knowledge of characteristics (e.g., number, locations, and transmitting powers) of transmitters (base stations), which is strong and may not hold in practice, especially for low-altitude economy applications that operate in unlicensed spectra like WiFi, where there often exist many transmitters unknown to users.
In contrast, the sampling-based problem uses certain prior information, e.g., terrain features, obstacle layouts, and limited radio samples collected by sensors sparsely distributed in the geographical region of interest, to predict radio maps.
In this paper, we focus on the sampling-based radio map construction problem.

\subsection{Related Work}
\smallskip

\noindent {\bf 2D Radio Map Construction\/} \ \ \
There are many existing works in the literature that construct a radio map for a 2D geographical region from radio samples collected by sensors distributed in that 2D region.
Their approaches can generally be classified into two categories: interpolation-based approaches and deep learning-based approaches.
Interpolation-based approaches, e.g., radial basis functions (RBF)~\cite{rbf}, spline~\cite{spline}, and ordinary kriging~\cite{kriging}, employ mathematical methodologies to construct radio maps.
Although theoretically sound, they often fail to produce high-quality maps for dynamic and complex real-world environments.
This is because they rely on mathematical calculations and are unaware of certain critical prior information, such as terrain features and obstacle layouts, which can significantly impact the propagation of radio signals.

Recent advances in radio map construction have demonstrated superior performance of deep learning techniques over conventional interpolation-based methods.
For example, Levie \emph{et al.} \cite{RadioUNet} proposed a UNet-based scheme,
Teganya \emph{et al.} \cite{AE} designed an autoencoder-based method,
He \emph{et al.} \cite{ResNet} designed a ResNet-based approach, and
Liu \emph{et al.} \cite{Liu25:ArXiv:diffusion} developed a diffusion model-based framework.
Other deep learning-based approaches include Feedforward Neural Networks~\cite{FEED_ref}, Generative Adversarial Networks  (GANs)~\cite{GAN_ref}, and Graph Attention Networks~\cite{RadioGAT}.

\smallskip

\noindent {\bf 3D Radio Map Construction\/} \ \ \
The problem of 3D radio map construction is to construct a 2D radio map at any desired altitude (target height) within a 3D region of interest, using radio samples collected by sensors distributed at different heights throughout that 3D region.
In contrast to 2D radio map construction, there are only a few existing works that study 3D radio map construction in the literature.
Hu \emph{et al.}~\cite{Hu23:TVT:3DMap} developed a GAN-based 3D construction framework, but required complete transmitter information (number, locations, and transmitting powers) in addition to radio samples.
Zhao \emph{et al.}~\cite{zhao20253d} proposed a diffusion model approach, but assumed the presence of only a single base station within the 3D region and required prior knowledge of the base station’s location in addition to radio samples. 
Different from~\cite{Hu23:TVT:3DMap,zhao20253d} which required the knowledge of transmitters' (base stations') characteristics for 3D radio map construction, Krijestorac \emph{et al.}~\cite{krijestorac2021spatial} leveraged UNet to estimate maps only from radio samples, and Chen \emph{et al.}~\cite{Chen15:Letter:3DMap} employed the convolutional autoencoder to estimate maps only from radio samples.
Notably, Zhang \emph{et al.}~\cite{Zhang24:TCCN:3DMap} introduced a large 3D radio map dataset \emph{SpectrumNet}, which is based on ray tracing simulations and facilitates the training and evaluation of 3D radio map construction models.

\subsection{Our Contribution}

In this paper, we study the fine-grained 3D radio map construction problem with an objective of constructing high-resolution 2D radio maps at arbitrary altitudes (heights) within a 3D region of interest, using radio samples collected at different heights throughout that 3D region.
This problem is uniquely challenging due to \emph{ultra-sparse sampling}, where the high-resolution requirement of radio maps conflicts with the ultra-small number of radio samples collected.

{Nowadays, more and more low-flying aircrafts like drones can carry sensors and collect radio samples within a 3D space.
Due to the real-time requirement of radio resource management (e.g., IEEE standard 802.22 \cite{IEEE80222} requires that the operating channel used for communication between a base station and Customer Premise Equipments (CPEs) within a Wireless Regional Area Network (WRAN) cell shall be sensed at least every 2 seconds for the signal types as required by a particular regulatory domain), it is impractical to use a single drone to densely sample a large 3D space within several seconds.
Considering that the cost of sensors that can collect radio samples is very high (e.g., even a low-end sensor like the Ettus USRP B200mini can cost over $1,000$ USD~\cite{EttusB200mini}), it is practically significant for fine-grained 3D radio map construction approaches to be able to accurately construct radio maps from a minimal number of radio samples collected by a few drones.
In this paper, we aim to construct fine-grained 3D radio maps from ultra-small sampling rates (e.g., $\le 0.1\%$ sampling rates).}
In contrast, existing 3D radio map construction approaches, such as those proposed by~\cite{Hu23:TVT:3DMap,zhao20253d,krijestorac2021spatial,Chen15:Letter:3DMap}, do not take into account the ultra-sparse sampling challenge and typically require a sampling rate of at least $3\%$ to predict high-quality radio maps.
Their performance under ultra-small sampling rates remains an open question.

\begin{figure*}[]
	\centering
	\includegraphics[width=0.8\linewidth]{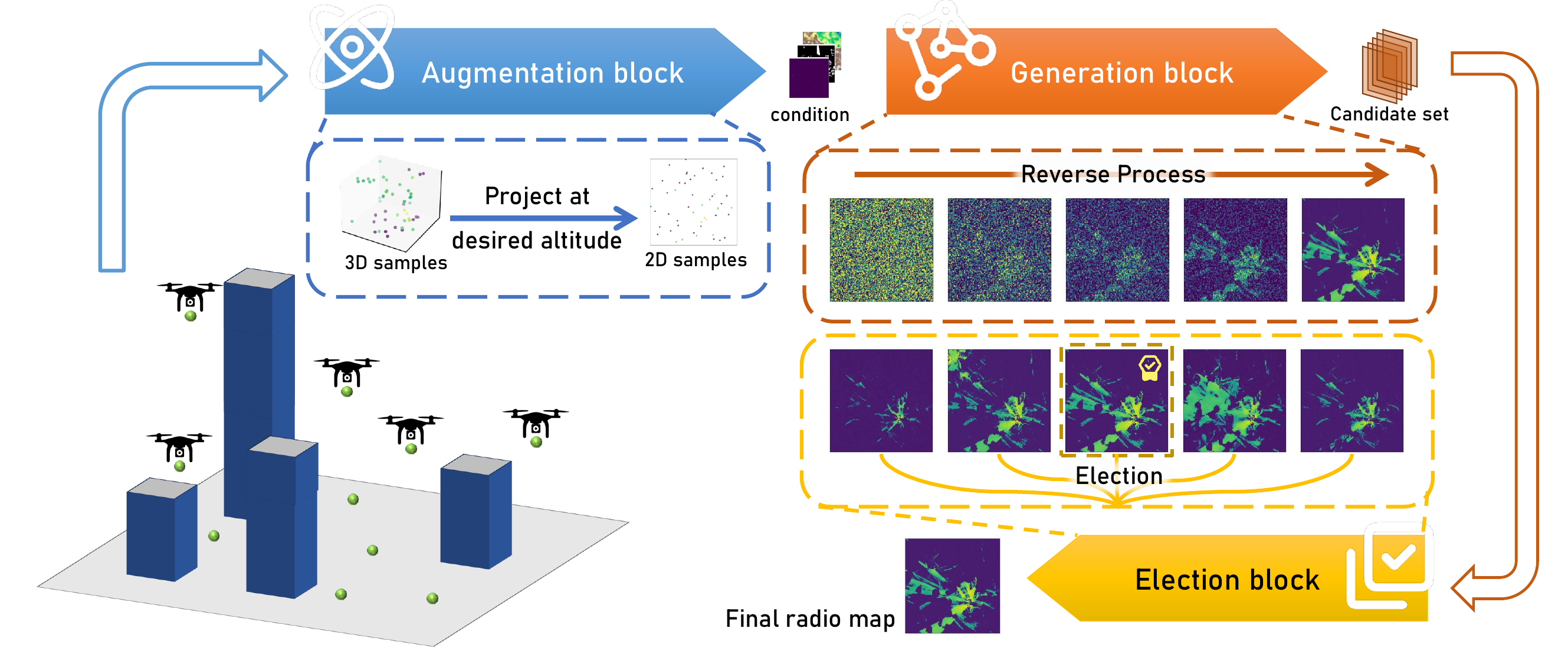}
    \caption{{RadioLAM -- a fine-grained 3D radio map construction paradigm built on generative LAMs.}}
	\label{diffMethod}
\end{figure*}

In order to deal with the ultra-sparse sampling challenge, we leverage the advanced Artificial Intelligence (AI) technique of generative Large AI Model (LAM) and propose RadioLAM for 3D radio map construction:
\begin{itemize}
    \item Creative power: Unlike traditional discriminative AI, LAM employs generative AI techniques and exhibits ``out-of-the-box thinking'' abilities to produce diverse outputs from limited inputs.
    For the problem of 3D radio map construction, we exploit this property to generate a candidate set of diverse fine-grained radio maps for the target 2D area with ultra-low sampling rates.
    Although most of the maps in the candidate set may be inaccurate, we expect that the best map in the set is accurate, enabled by the creative power of LAM.

    \item Generalization capability: Unlike traditional small-scale AI model, LAM has a strong generalization capability of producing high-quality outputs even when presented with inputs that deviate from the training distribution. 
    For the problem of 3D radio map construction, we leverage this property to make high-quality radio map constructions for unseen 3D environments not represented in the training dataset, and also to enhance the performance for 3D environments similar to the training dataset.  
\end{itemize} 

As shown in Fig.~\ref{diffMethod}, RadioLAM consists of three key blocks: an 
 \emph{augmentation block}, a \emph{generation block}, and an \emph{election block}.
In the augmentation block, RadioLAM applies the principle of radio propagation to spatially project the radio samples collected at different altitudes (heights) onto the target 2D plane (at the desired altitude), incorporating environmental priors such as terrain features and building layouts to enhance projection accuracy.
In the generation block, RadioLAM employs an LAM to generate a candidate set of fine-grained radio maps for the target 2D plane, based on the radio projections which are the outputs of the augmentation block.
RadioLAM uses a Mixture of Experts (MoE) architecture for the design of LAM to efficiently enhance the generative performance of LAM across diverse 3D environments. 
Because of the ultra-small sampling rate, the majority of the maps in the candidate may be inaccurate, with only a minority being accurate. 
Then in the election block, RadioLAM selects the best map from the candidate set, by using the radio propagation principle as a guide to identify the map demonstrating the closest adherence to the principle.
Besides, in the election block, RadioLAM also leverages a Test-Time Adaptive (TTA) scheme to dynamically adjust the generative noise level of LAM during inference, to maintain an effective balance between output diversity and physical plausibility of LAM.

{In the literature, there are already some existing works that take advantage of a generative model's capability of generating diverse outputs as well as a post-processing selection stage to solve various complex problems.
For example, DATID-3D~\cite{DATID-3D}, a diffusion-based model for text-to-image generation, produces diverse outputs and then employs CLIP-based scoring and pose reconstruction to filter for the best result.
TopoDiff~\cite{topodiff}, a diffusion-based model for protein backbone generation, creates a diverse set of protein structures and subsequently selects the most viable one based on biophysical rules.
The Self-Consistency method~\cite{Self-Consistency} for large language models samples multiple reasoning paths from a decoder, marginalizes specific reasoning steps, and aggregates the final answers to select the most frequent one.}

The main contributions of this paper are summarized as follows: 

\begin{itemize}
\item We study the fine-grained 3D radio map construction problem which aims to construct high-resolution radio maps for target 2D areas at desired altitudes (target heights) within a 3D space of interest, using radio samples collected at different heights throughout that 3D region.
This problem is challenging because the number of collected radio samples is far fewer than the high-resolution requirement of the radio map to be estimated.
We develop RadioLAM -- a {novel fine-grained 3D radio map construction paradigm} which addresses this challenge by the creative power and the strong generalization capability of generative LAM.
To the best of our knowledge, we are the first to take into account the challenge of ultra-sparse sampling when designing algorithms to solve the 3D radio map construction problem.

\item RadioLAM consists of three key blocks: 1) an augmentation block, which leverages the principle of radio propagation to spatially project the multi-altitude radio samples collected by sensors onto the target 2D area at the desired altitude;
2) a generation block, which employs a diffusion-based LAM under the MoE architecture to generate a candidate set of fine-grained radio maps for the target 2D area, based on the radio projections from the augmentation block;
3) an election block, which uses the radio propagation principle as a guide to find the best map from the diverse candidates generated by the generation block, and utilizes a TTA scheme to dynamically adjust the noise level of LAM during inference.
With the three blocks, RadioLAM is able to obtain high-quality solutions for the fine-grained 3D radio map construction problem from ultra-sparse sampling.

\item Extensive simulations based on the open-source dataset \emph{SpectrumNet} show that RadioLAM is effective for constructing fine-grained radio maps across diverse 3D radio propagation environments at different target altitudes.
Specifically, simulations show that for different environments, including rural, suburban, ordinary urban, and dense urban, and for different target altitudes, 
1) RadioLAM produces high-quality, fine-grained radio maps from an ultra-small sampling rate of $0.1\%$;
2) radio maps generated by RadioLAM are over $50\%$ better than those generated by state-of-the-art (SOTA) under the $0.1\%$ sampling rate;
3) RadioLAM requires at most one fourth of the sampling rate needed by SOTA to construct radio maps with comparable qualities.
{Furthermore, experiments based on the open-source real-world dataset \emph{AERPAW} also demonstrate that the radio maps achieved by RadioLAM are much more accurate than those schieved by SOTA.}
\end{itemize}

The rest of the paper is organized as follows. 
In Section~\ref{system-model-and-problem-description}, we describe the system model and introduce the 3D radio map construction problem. 
In Section~\ref{scheme}, we present the key ideas of the design of RadioLAM. 
In Section~\ref{scheme-detail}, we offer the design details of RadioLAM. 
In Section~\ref{Experimental}, we perform extensive simulations {and real-world experiments} to evaluate RadioLAM.
In Section~\ref{Conclusion}, we conclude the paper.

\section{Problem Description} \label{system-model-and-problem-description}

\begin{table}[]
	\centering
	\caption{Summary of key notations.}
	\renewcommand{\arraystretch}{1.3}
	\begin{tabular}{cp{2.2in}}
		\hline
		Symbol & Definition \\ \hline
        $\mathcal{A}$& 3D area of interest\\
        $k$& number of sensors collecting radio samples at different heights within $\mathcal{A}$\\
        $\mathcal{S}$& set of $k$ sensors that collect radio samples at different heights within $\mathcal{A}$\\
        $X$& length of the discrete $\mathcal{A}$\\
        $Y$& width of the discrete $\mathcal{A}$\\
        $H$& height of the discrete $\mathcal{A}$\\
         $\mathbf{S}_k$& an $X\times Y\times H$ matrix, where the element at index $(x,y,h)$ is 0 if there is no sensor located at that grid point; otherwise, it is the RSS value of the sample collected by the sensor\\
   $\mathbf{B}$& layout of buildings and an $X\times Y\times H$ matrix, where the element at index $(x,y,h)$ is $1$ if certain building is located at that grid point; otherwise, it is $0$\\
   $\mathbf{T}$& feature of terrain and an $X\times Y$ matrix, where the element at index $(x,y)$ specifies the local ground altitude at the horizontal position $(x,y)$ \\
   $h_t$ &  the desired altitude (target height) of the 2D plane where a high-resolution radio map is required to be estimated from $\mathbf{S}_k$, $\mathbf{B}$, and $\mathbf{T}$\\
   \hline
\end{tabular}
\label{Notations}
\end{table}

Let $\mathcal{A}\subset\mathbb{R}^3$ represent a 3D area of interest. 
Within this space, consider a set of sensors $\mathcal{S}=\{1,2,\cdots,k\}$ (the number of sensors is $k$) where each sensor $s\in\mathcal{S}$ is located at a unique 3D coordinate $a_s\in\mathcal{A}$.
Each sensor $s$ collects a radio sample that measures the RSS $r_s(f)$ in a frequency band $f\in\mathcal{F}$ at the location $a_s$.
These radio samples $r_s(f)$ ($s\in\mathcal{S}$ and $f\in\mathcal{F}$) are reported to a fusion center (e.g., a base station or a cloud server).
To facilitate analysis, in this paper, we restrict our consideration to a single frequency $f$, considering only spatial samples with RSS $r_s$ for each $s\in\{1,2,\cdots,k\}$ at the location $a_s\in\mathcal{A}$.

We discretize the continuous spatial area $\mathcal{A}$ into a structured grid of dimension $X\times Y\times H$, where each sensor location $a_s$ maps to a unique grid index $(x_s,y_s,h_s)$.
Here $x_s\in\{0,1,\cdots,X-1\}$ and $y_s\in\{0,1,\cdots,Y-1\}$ represent horizontal coordinates, while $h_s\in\{0,1,\cdots,H-1\}$ denotes the vertical altitude (height) index.
Define $\mathbf{S}_k$ as an $X\times Y\times H$ matrix, where its element at index $(x,y,h)$, i.e., $\mathbf{S}_k(x,y,h)$, is 0 if there is no sensor located at that grid point; otherwise, it is the RSS value of the sample collected by the sensor located at $(x,y,h)$.  

To establish realistic environmental modeling, we take into account environmental obstacles, specifically buildings and terrain characteristics.
The layout of buildings is represented by an $X\times Y\times H$ binary matrix $\mathbf{B}$, where its element at index $(x,y,h)$, i.e., $\mathbf{B}(x,y,h)$, is $1$ if certain building is located at that grid point; otherwise, it is $0$.
The characteristic of terrain is represented by an $X\times Y$ elevation matrix $\mathbf{T}$, where its element at index $(x,y)$, i.e., $\mathbf{T}(x,y)$, specifies the local ground altitude at the horizontal position $(x,y)$.
Both buildings and terrain span the entire 3D area $\mathcal{A}$, potentially occupying multiple adjacent grid points.
We consider the occupied grid points to have no RSS; however, they do affect RSS in surrounding areas by reflecting radio waves back into the environment and by blocking or attenuating radio waves passing through them.

\begin{figure}[]
	\centering
	\includegraphics[width=\linewidth]{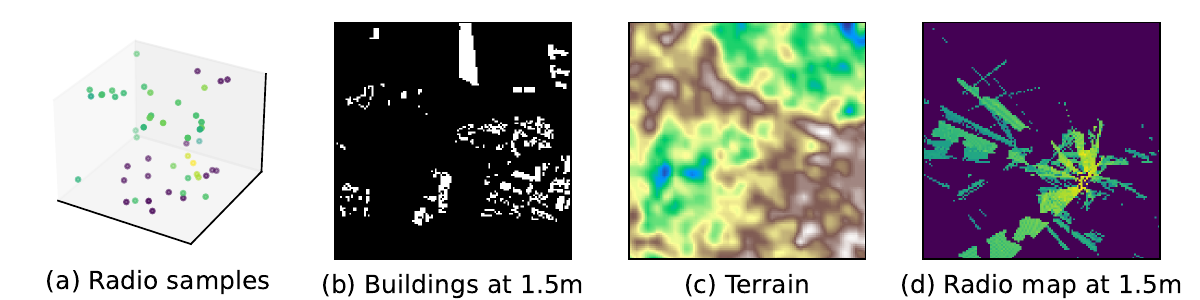}
	\caption{An example of the fine-grained 3D radio map construction problem, with an objective of using (a), (b), and (c) to produce (d).}
	\label{fig:input}
\end{figure}

Given a desired altitude (target height) $h_t\in\{0,1,\cdots,H-1\}$, the problem of fine-grained 3D radio map construction requires the fusion center to use $\mathbf{S}_k$, $\mathbf{B}$, and $\mathbf{T}$ to infer (estimate) the RSS values at grid points $(x,y,h_t)$ for all $x\in\{0,1,\cdots,X-1\}$ and all $y\in\{0,1,\cdots,Y-1\}$.
Fig.~\ref{fig:input} shows an example of the fine-grained 3D radio map construction problem.
Fig.~\ref{fig:input}a describes radio samples collected within the 3D area $\mathcal{A}$.
Fig.~\ref{fig:input}b shows layout of buildings for the target 2D plane at height $1.5$m.
Fig.~\ref{fig:input}c illustrates terrain features.
Fig.~\ref{fig:input}d gives an estimated fine-grained radio map for the target 2D plane at height $1.5$m.
In this example, the problem of 3D radio map construction requires to use Fig.~\ref{fig:input}a, Fig.~\ref{fig:input}b, and Fig.~\ref{fig:input}c to predict Fig.~\ref{fig:input}d.

\section{RadioLAM: Basic Ideas}\label{scheme}

\begin{figure*}[]
	\centering
	\includegraphics[width=0.8\linewidth]{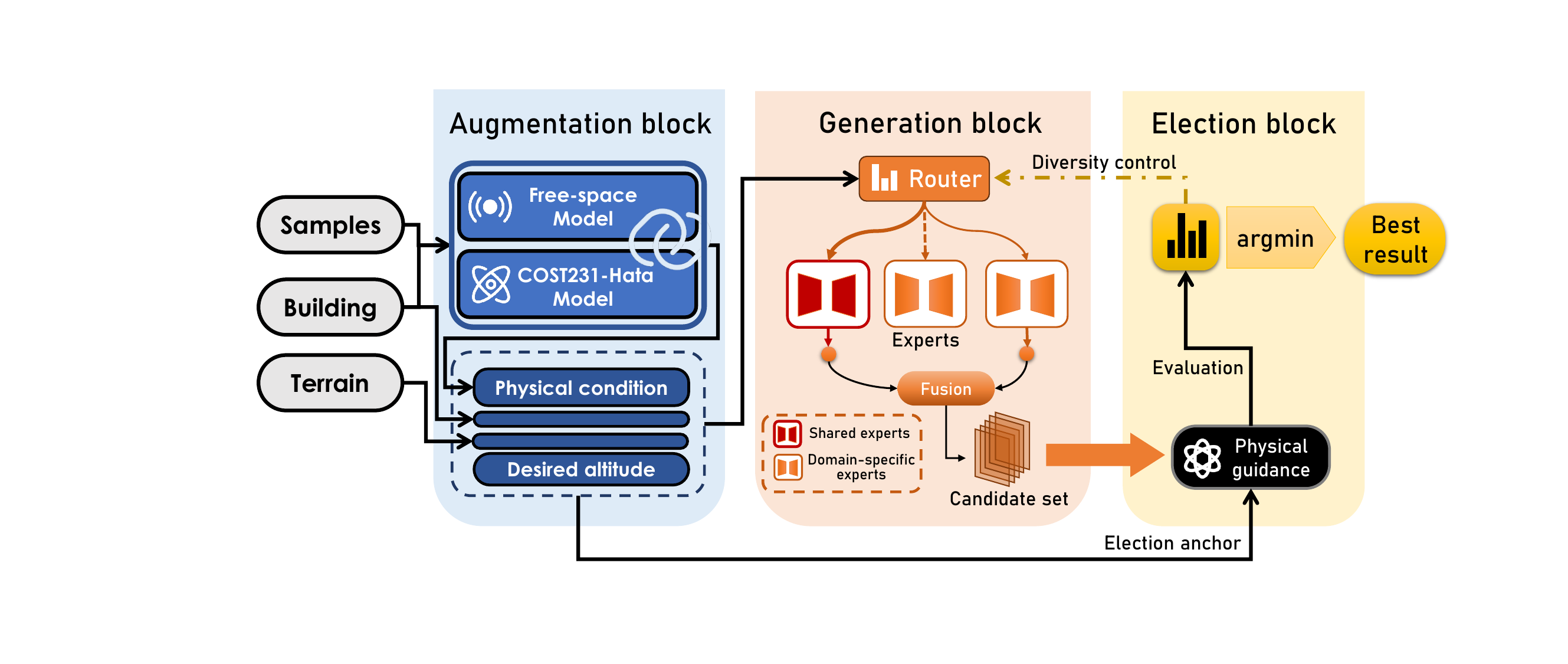}
	\caption{RadioLAM -- an architecture overview.}
	\label{overview}
\end{figure*}

The problem of fine-grained 3D radio map construction is very challenging due to the sparsity of collected radio samples relative to the high-resolution requirement for the radio map to be estimated.
To overcome this challenge, we propose RadioLAM.
As shown in Fig.~\ref{overview}, RadioLAM consists of three key blocks: an augmentation block, a generation block, and an election block.
In the following, we briefly describe functions of these three blocks, respectively.

\smallskip

\noindent {\bf Block 1: Augmentation block\/} \ \ \
The 3D radio map construction is challenging due to that the number of radio samples, i.e., $k$, is very small in practice.
Furthermore, the number of radio samples collected at the target height $h_t$ can be significantly smaller than $k$ or possibly $0$.
Therefore, it is impossible to estimate the radio map for the target 2D plane at height $h_t$ only from the coplanar radio samples.
To improve the qualities of fine-grained radio maps generated in the subsequent generation block for the target 2D plane, in this block, we leverage the principle of radio propagation to intelligently project the set of $k$ multi-altitude radio samples onto the target 2D plane at height $h_t$, with environmental parameters such as building layouts and terrain features taken into account.
The output of this process produces enhanced RSS estimates across the target 2D plane, effectively creating a denser set of virtual RSS measurements than could be obtained through direct sampling at the target height $h_t$ alone. 
These projected RSS values then serve as inputs of the subsequent generation block.

\smallskip

\noindent {\bf Block 2: Generation block\/} \ \ \
In this block, we use a diffusion-based LAM to produce a candidate set of diverse fine-grained radio maps for the target 2D plane, from the projected RSS values at different locations in that plane.
LAM has a ``out-of-the-box thinking" ability, enabling diverse solution generation from sparse inputs.
Although many candidates may be inaccurate, we expect that the solution space typically contains a few accurate solutions.
Moreover, LAM has a strong generalization capability, enabling robust performance even with input distributions that deviate from the training scenarios.
We leverage an MoE architecture for the design of the LAM, with each expert in the MoE being a radio map generator specializing in distinct propagation environments.
With the MoE architecture, the performance of LAM for generating high-quality radio maps for different kinds of radio propagation environments can be significantly improved.

\smallskip

\noindent {\bf Block 3: Election block\/} \ \ \
While the generation block demonstrates strong creative capabilities, a large proportion of its generated radio map candidates may be of low quality. 
It is necessary to design an election block following the generation block to identify the best map in the candidate set.
Specifically, in this block, we evaluate and select the most physically plausible solution from the set of candidates, by utilizing the radio propagation principle as a guide to identify the map that most closely adheres to the principle.
Moreover, in this block, we implement a TTA scheme to dynamically adjust the noise level of LAM during inference.
The objective is to maintain an effective balance between the output diversity and the physical plausibility of LAM, enhancing the robustness of RadioLAM to different input distributions during inference.

\section{RadioLAM: Design Details}\label{scheme-detail}

In this section, we describe design details of RadioLAM.

\subsection{Augmentation Block}

The problem of 3D fine-grained radio map construction aims to construct a high-resolution radio map for a target 2D plane at a desired altitude $h_t$, using radio samples collected within the 3D area of interest $\mathcal{A}$.
Since radio samples can be collected at altitudes different from $h_t$, it is impossible to construct a radio map for the target 2D plane only using samples collected at altitude $h_t$.
To address this challenge, the augmentation block projects the radio samples collected at altitudes different from $h_t$ onto the target 2D plane, i.e., predicts the RSS value located at $(x_s,y_s,h_t)$ from the sample $s$ collected at the location $(x_s,y_s,h_s)$ for each $s\in\mathcal{S}$ with $h_s\neq h_t$.
This projection significantly increases the number of known or estimated RSS values on the target plane compared to using only samples collected at altitude $h_t$.
The augmented dataset thereby enables the subsequent generation block to construct high-quality radio maps for the target plane.

For each $s\in\mathcal{S}$ with $h_s\neq h_t$, to predict the RSS value located at $(x_s,y_s,h_t)$ from that located at $(x_s,y_s,h_s)$, we design an algorithm that combines the free-space propagation model~\cite{Friis} and the Hata propagation model~\cite{Hata}.

\smallskip

\noindent {\bf Free-space Propagation Model\/} \ \ \
From Friis' free-space propagation formula~\cite{Friis}, we have
\begin{equation}
    P_r(d) = P_t\frac{G_tG_r\lambda^2}{(4\pi)^2 d^nL},
\end{equation}
where $P_t$ is the power fed into the transmitting antenna, $P_r$ is the power available at the output terminal of the receiving antenna, $G_r$ is the effective area of the receiving antenna, $G_t$ is the effective area of the transmitting antenna, $d$ is the distance between antennas, $\lambda$ is the wavelength, $L$ is system loss, and $n$ is path loss exponent.

Consider $\mathcal{T}=\{T_1,T_2,\cdots,T_N\}$ as the set of transmitters, and each transmitter $\tau\in\mathcal{T}$ is located at $(x_{\tau},y_{\tau},h_{\tau})$.
From the Friis' free-space propagation formula, the received power of the point $p$ located at $(x_p,y_p,h_p)$ can be calculated as:
\begin{equation}
\begin{aligned}
    P_{\text{free}}(p) =& \sum_{\tau\in\mathcal{T}} \biggl[K_{\tau}\cdot \left((x_p-x_{\tau})^2\right.\\&\left.+(y_p-y_{\tau})^2+(h_p-h_{\tau})^2\right)^{-\frac{n}{2}}\biggr].\label{eqn:free}
\end{aligned}
\end{equation}
where
\begin{equation}
K_{\tau}=\frac{P_{\tau}G_{\tau}G_r\lambda^2}{(4\pi)^2 L}.
\end{equation}

To employ the equation~\eqref{eqn:free} to calculate RSS values, we need to know the location of each transmitter $\tau$ and $K_{\tau}$.
We note that locations of transmitters are not the input to our problem.
In the augmentation, (i) we use the interpolation-based method RBF~\cite{rbf} to estimate locations of transmitters from collected radio samples; and (ii) use the Levenberg-Marquardt algorithm~\cite{LM1963} to estimate $K_{\tau}$.
{Although the estimation of transmitter locations as well as $K_{\tau}$ introduces unavoidable errors to the augmentation block and subsequent to the generation block of RadioLAM, as evaluated comprehensively later in Section~\ref{Experimental} (especially in Table~\ref{tab:maintable} and Table~\ref{tab:ab-table}), the radio maps generated by RadioLAM are much better than SOTA, and the augmentation block contributes significantly to the superior performance of RadioLAM (the radio maps generated by RadioLAM becomes much worse if the augmentation block is removed).}

The RSS value located at $(x_s,y_s,h_t)$ can be accurately predicted by the free-space propagation model when an unobstructed line-of-sight (LOS) path exists between $(x_s,y_s,h_t)$ and locations of transmitters.
However, in environments with obstructions such as buildings or terrain that block radio wave propagation, the free-space model yields inaccurate predictions.
For non-line-of-sight (NLOS) conditions, an alternative propagation model must be employed to predict RSS values.

\smallskip

\noindent {\bf Hata Propagation Model\/} \ \ \
We leverage the COST231-Hata model for RSS prediction in NLOS setting.
Specifically, we can use Hata model to predict the RSS value located at $(x_s,y_s,h_t)$ from the RSS value of the radio sample collected at $(x_s,y_s,h_s)$. 
According to the Hata model, we have
\begin{equation}\label{eqn:Hata}
    PL = \beta - a(h),
\end{equation}
where $PL$ is path loss, and the value of $\beta$ depends on many parameters including the frequency band, information of transmitters, and distance from transmitters.
However, the value of $\beta$ remains the same for the location $(x_s,y_s,h_t)$ and $(x_s,y_s,h_s)$.
We only need to calculate $a(h_t)$ and $a(h_s)$ in order to predict the RSS at $(x_s,y_s,h_t)$ from that at $(x_s,y_s,h_s)$:
\begin{equation}
    P_{\text{Hata}}(p)=P_{\text{Hata}}(\hat{p})+a(h_{\hat{p}})-a(h_p),
\end{equation}
where the points $p$ and $\hat{p}$ share the same horizontal position, i.e., $x_p=x_{\hat{p}}$ and $y_p=y_{\hat{p}}$.
$a(\cdot)$ is a correction factor for mobile antenna height.
The specific function of $a(\cdot)$ can be found in~\cite{Hata}.

{Overall, the free-space propagation model is accurate for RSS predictions of LOS locations, but produces large errors for RSS predictions of NLOS locations.
The Hata model is effective for RSS predictions along the vertical lines of the collected radio samples, bur fails to predict RSS values of other locations as the parameter $\beta$ in equation~\eqref{eqn:Hata} is unknown and hard to estimate.
Given these inherent limitations of the two propagation models, as well as that even the information of LOS and NLOS is unknown, the augmentation block functions as a ``smart initializer".
It heuristically combines the two models to generate physically-informed but incomplete estimates only for a few locations where the two propagation models are most reliable.
The subsequent generation block, powered by a diffusion-based LAM, is then essential to infer the RSS values for the vast majority of locations where the propagation models fail.
}

\begin{figure}[]
	\centering
	\includegraphics[width=0.8\linewidth]{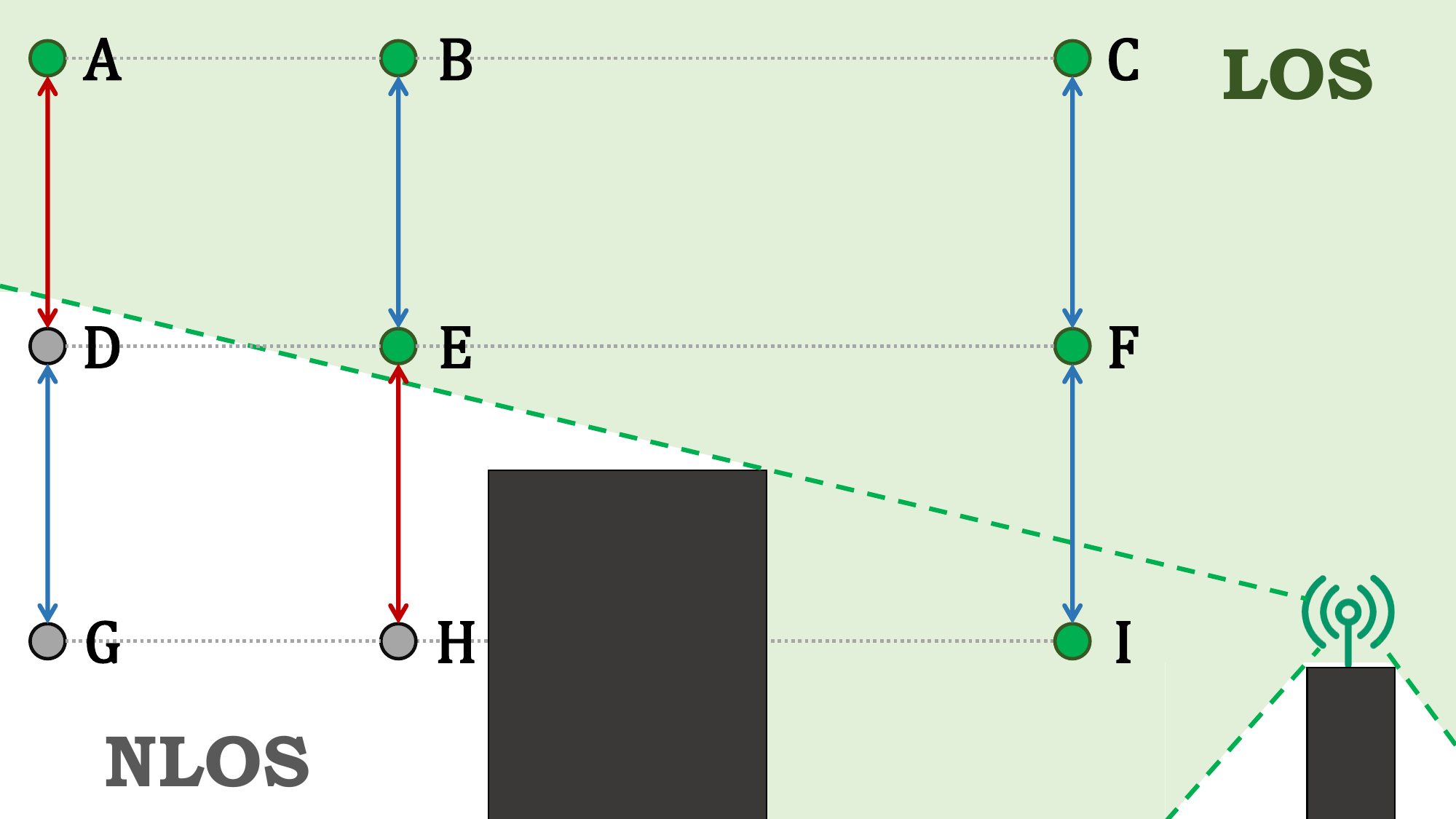}
	\caption{An illustration of LOS and NLOS locations.}
	\label{fig:phy2}
\end{figure}

Specifically, the two propagation models are combined as follows:
\begin{equation}
    P_{\text{aug}}(p)  = (1-w)\cdot P_{\text{free}}(p) +w\cdot P_{\text{Hata}}(p),\label{eqn:aug} 
\end{equation}
where $w\in(0,1)$ is a weight factor.
From Fig. \ref{fig:phy2}, we observe that LOS locations are typically located at high altitudes, while NLOS locations are typically located at low altitudes.
Motivated by this observation, we use the following definition for $w$:
\begin{equation}
    w=2^{-\frac{h_p}{U}},
\end{equation}
where $U$ is a positive hyperparameter.
For $p$ located at a high altitude, $w$ will be small and hence the free-space propagation model will play a more important role in RSS prediction compared to the Hata model;
while for $p$ located at a low altitude, $w$ will be large and hence the Hata model will play a more important role in RSS prediction compared to the free-space propagation model.

Moreover, it is intuitive that~\eqref{eqn:aug} yields higher prediction accuracy for RSS at $(x_s,y_s,h_t)$, when both $(x_s,y_s,h_t)$ and $(x_s,y_s,h_s)$ share the same propagation condition (either both LOS or both NLOS locations), compared to cases where their propagation conditions differ.
For example, in Fig. \ref{fig:phy2}, we prefer predicting $D$ from $G$ rather than from $A$; similarly, we prefer predicting $E$ from $B$ rather than from $H$.
To optimize the augmentation process, motivated by the above intuition, we add a filtering strategy: considering that the RSS value which is extremely small is more likely to be located at the NLOS region, given a small predefined threshold $\theta$, we will not use the radio sample $s$ for prediction if its RSS value $r_s<\theta$, and even when we use $s$ for prediction and obtain a predicted RSS $\hat{r}_s$, we will drop $\hat{r}_s$ if $\hat{r}_s<\theta$.

\subsection{Generation Block}

The objective of this block is to generate a diverse set of fine-grained radio map candidates for the target 2D plane at altitude $h_t$.
We design an LAM under a diffusion-based MoE architecture to achieve this goal.

\begin{figure*}[]
	\centering
	\includegraphics[width=0.8\linewidth]{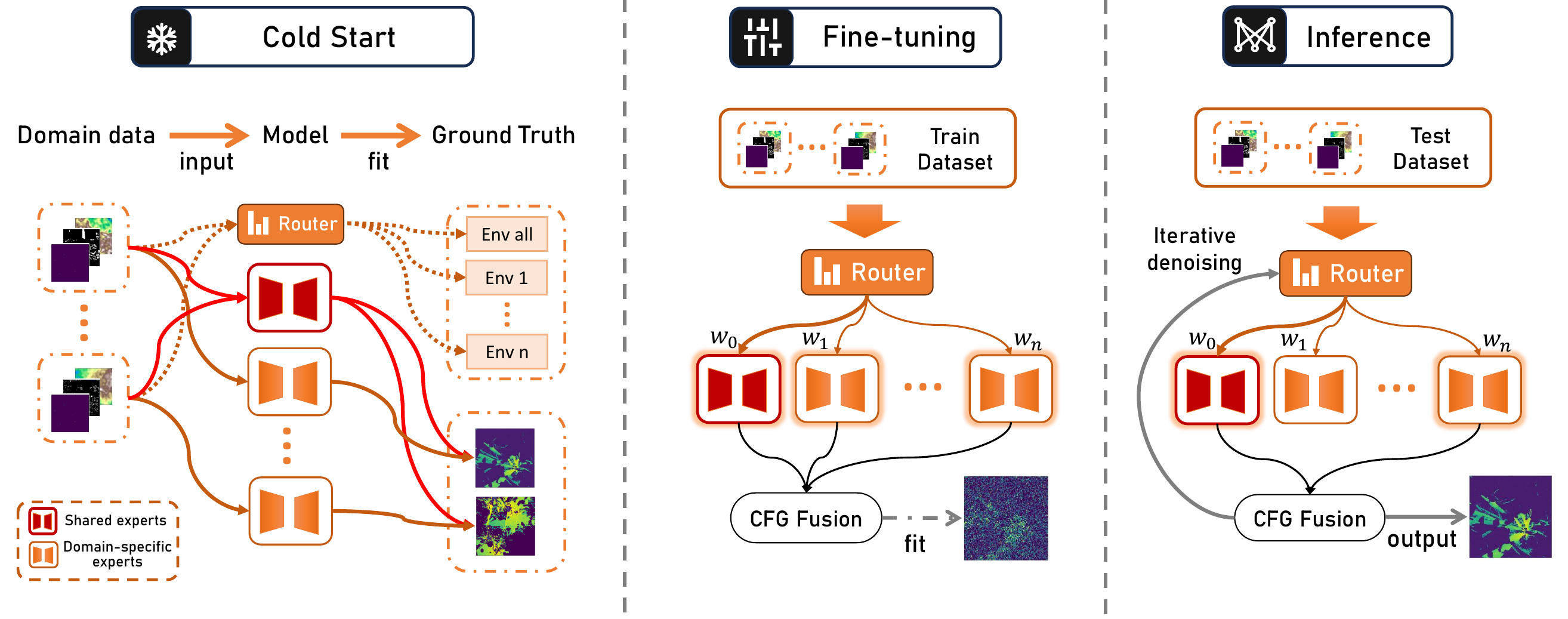}
	\caption{The generation block employs an MoE-based LAM to generate diverse radio map candidates.}
	\label{fig:generation}
\end{figure*}

We note that in practice, radio propagation characteristics vary significantly across different environments.
For example, as shown later in Fig.~\ref{fig:case-study-o} and Fig.~\ref{fig:case-study-r} of the evaluation section, urban and rural environments exhibit different radio map patterns.
To enhance the generation ability of this block for varying or even unseen propagation characteristics, we use an MoE architecture for the design of LAM.
The LAM itself has a strong generalization capability of generating high-quality radio maps for diverse environments.
In addition, the MoE architecture incorporates domain-specific experts, each optimized for a particular environmental condition.
Compared to using only a monolithic model for generation, the MoE-based model enables better radio map generation for different environments. 

Specifically, inspired by the DeepSeekMoE~\cite{DeepSeekMoE} architecture, our diffusion-based LAM consists of a shared expert (generator) handling fundamental propagation patterns, multiple domain-specific experts (generators) specializing in distinct environmental conditions, and a router dynamically selecting appropriate experts by generating different weights for them, as shown in Fig.~\ref{fig:generation}.
The training of this LAM employs a two-phase approach: a cold start phase and a fine-tuning phase:
\begin{itemize}
    \item Cold start of training: The objective of this phase is to utilize available dataset to train the shared expert, domain-specific experts, and router, separately.
    The shared expert has a UNet structure and will be trained by the complete radio map dataset, complete building layout dataset, and complete terrain feature dataset.
    Each domain-specific expert also has a UNet structure but will be trained only by the radio map data, building layout data, and terrain feature data corresponding to the specialized domain (environment).
    For all the experts, we employ the training method of Denoising Diffusion Probabilistic Model (DDPM)~\cite{DDPM}.
    The router has a ResNet18 structure and will be trained by the complete building layout dataset and complete terrain feature dataset, both of which are pre-labeled.    

    \item Fine-tuning of training: The objective of this phase is to update the parameters of the shared expert, domain-specific experts, and router simultaneously by training on the complete dataset.
    Let all experts and router be activated in this phase.
    For the complete dataset including radio maps, building layouts, and terrain features, in each instance, the router will first generate different weights for different experts.
    Then, each expert will generate radio maps based on the radio samples, building layout, and terrain feature.
    Finally, the generated radio maps from all experts will be fused using the Classifier-Free Guidance (CFG)~\cite{CFG} as the fusion method based on the weights from the router.
\end{itemize}

During inference, the MoE-based LAM leverages a Denoising Diffusion
Implicit Model (DDIM)~\cite{DDIM} procedure to generate radio map candidates.
DDIM is an iterative procedure.
In each iteration, it consists of a forward process (diffusion process) which adds noise to radio maps and a reverse process (inverse diffusion process) which generates radio maps through denoising.
In each iteration, the radio samples, building layout, and terrain feature will be the inputs to LAM.
For LAM, first, the router will generate different weights for different experts based on the building layout and terrain feature.
Then, only the shared expert and the domain-specific expert with the largest weight will be activated to generate radio maps based on all inputs.
Next, radio maps generated by the activated experts will be fused using CFG.
Compared to DDPM, DDIM significantly reduces the inference time while preserving the high quality of the generated results. 
Besides, DDIM can be seamlessly applied after DDPM-based model training, allowing for a straightforward conversion of a DDPM model to a DDIM model without the need for retraining.
At the end of the iterative DDIM-based inference, LAM will produce a candidate set of diverse fine-grained radio maps.

Here we highlight that the training objective of our diffusion model-based generation block is to minimize the Mean Square Error (MSE) defined later in Equation~\eqref{eqn:mse}, similar to many traditional deep learning approaches.
As shown in some recent studies \cite{TimeDiff,ARMD,MADM}, the diffusion model could achieve lower MSE compared with traditional deep learning models for certain MSE-minimizing problems. 
Especially, the theoretical work in~\cite{bestMSE} designed a denoiser for the diffusion model and proved that the solution of such a diffusion model is very close to the MSE-minimal solution with polynomial-time convergence guarantee under mild assumptions.
Those works motivate the diffusion-based design of RadioLAM.

During inference, the diffusion-based generation block iteratively denoises an initial random noise conditioned on the available inputs: RSS measurements collected by sensors, building layouts, and terrain data. 
While each denoising step is guided by a UNet trained to minimize MSE, the different noise initializations may lead to different outputs.
This outcome variation, which stems from the generative formulation of the diffusion model, motivates our inclusion of an election block to identify the best output from multiple inferences of the generation block.

\subsection{Election Block}
The objective of this block is to identify the best map in the diverse set of fine-grained radio map candidates generated by the generation block.
We employ a radio propagation model (consistent with our augmentation block) as a guide to identify the candidate that demonstrates the highest conformance to it and consider such a candidate as the best map.
Moreover, we incorporate a TTA-based scheme in this block, to dynamically adjust the noise level of LAM, ensuring both the generation diversity and physical plausibility of LAM.

Considering that the true radio map (ground truth) is unknown during inference, it is impossible to directly find the best map from the candidate set based on a comparison to the true map.
Although the RSS value of each location in the target 2D plane at altitude $h_t$ (i.e., each location in the true map) is unknown, at the end of the augmentation block, we know RSS values of several locations in the 2D plane at altitude $h_t$.
These RSS values either come from radio samples collected at altitude $h_t$, or are predicted (projected) by radio samples collected at the same horizontal positions but at different altitudes using the radio propagation model~\eqref{eqn:aug}.
Intuitively, we can compare the radio map candidates with these RSS values to identify the best one.

Specifically, suppose $\bar{r}_s$ is the RSS value of the location $(x_s,y_s)$ in the generated radio map.
Suppose $\hat{r}_s$ is the RSS value predicted from the radio sample $s\in\mathcal{S}$ with $h_s\neq h_t$ in the augmentation block (let $\hat{r}_s=r_s$ if $h_s=h_t$, and let $\hat{r}_s=\bar{r}_s$ if we do not project $s$ onto the target 2D plane).
For each map in the candidate set, we calculate the following distance metric:
\begin{equation}
    D_{\text{ele}}~=~\sum_{s\in\mathcal{S}}(\bar{r}_s-\hat{r}_s)^2.
    \label{eqn:diversity}
\end{equation}
We consider the radio map that minimizes $D_{\text{ele}}$ among all the maps in the candidate set as the best map.

Note that RadioLAM addresses the ultra-sparse sampling challenge of 3D radio map construction using the creative power of LAM, by generating a diverse set of radio map candidates from ultra-small number of radio samples.
At the end of this election block, we design a control algorithm to maintain appropriate diversity of LAM, by adjusting LAM's noise level based on the variance of qualities of the radio map candidates.

Considering that for each radio map candidate, we can quantify its quality by calculating the distance metric~\eqref{eqn:diversity}.
Then, we quantify the diversity of radio map candidates by calculating the variance of corresponding distance metrics.
Let $\text{Var}$ be such a variance. 
We optimize the diversity of LAM by adjusting the zero-mean noise which we inject to the CFG weights during the CFG-based fusion of LAM:
\begin{equation}
    \sigma_{t+1}=\left\{\begin{array}{ll}
        \min(\sigma_{t} +\Delta\sigma,\sigma_{m}), & \text{Var}<V\\
        \sigma_{t}/2, & \text{Var}\ge V
        \end{array}\right.
\end{equation}
where $\sigma_{t}$ is the variance of the zero-mean noise at the current timestep $t$, $\sigma_{t+1}$ is the variance of the zero-mean noise at the next timestep $t+1$, $\Delta\sigma$ is a hyperparameter of increment step, $V$ is a hyperparameter of variance threshold, and $\sigma_{m}$ is a hyperparameter of noise threshold.
With this mechanism, the noise will increase if the current output diversity of LAM is relatively low; otherwise, the noise will reduce.
As a result, the output diversity of LAM is expected to remain at an appropriate level.

\section{Evaluation}\label{Experimental}

In this section, we evaluate RadioLAM.
{A comprehensive evaluation is first conducted using a large-scale, open-source dataset simulated by ray tracing. 
Subsequently, RadioLAM's effectiveness is also verified on a small-scale, open-source dataset collected in real world.}
{Specifically, based on the large-scale simulated dataset,} we first use a case study to evaluate RadioLAM's performance and demonstrate RadioLAM's capability of constructing fine-grained radio maps for target 2D planes across a 3D area of interest from ultra-sparse sampling.
Then we use extensive simulations to compare RadioLAM against various baseline algorithms, by estimating radio maps for different 2D planes at varying heights across diverse kinds of 3D environments.
Next, we conduct ablation studies to analyze the individual contributions of the augmentation block, the generation block, and the election block to the overall performance of RadioLAM.
Subsequently, we evaluate RadioLAM's performance under varying model sizes (i.e., different numbers of model parameters), {and use simulations to show that RadioLAM maintains significant performance advantages over baseline algorithms, even when baselines are scaled up to match or exceed RadioLAM's model size}.
{Finally, we use experiments based on the small-scale real-world dataset to demonstrate the superior performance of RadioLAM as compared to various baseline algorithms.}

\subsection{Simulation Settings}

We use the open-source dataset \emph{SpectrumNet}~\cite{Zhang24:TCCN:3DMap} for the evaluation.  
\emph{SpectrumNet} consists of radio maps in 5 different frequency bands: 150 MHz, 1.5 GHz, 1.7 GHz, 3.5 GHz and 22 GHz.
It includes 11 different kinds of geographical environment: dense urban, ordinary urban, suburban, rural, mountainous, forest, desert, grassland, island, ocean, and lake.
It takes into account the joint impact of terrain features and building layouts on radio propagation.
It uses OpenStreetMap~\cite{openstreetmap} to obtain real-world building layouts and terrain features of $15300$ different areas, each of which has a size of $1.28$ km $\times$ $1.28$ km in the horizontal plane, with a spatial resolution of $10$  m, resulting in radio maps of resolution $128\times 128$ (i.e., $X=Y=128$).
There may exist multiple transmitters in each area.
{It utilizes the ray tracing method of Matlab to simulate radio maps at $3$ different heights for each area: $1.5$ m, $30$ m, and $200$ m (i.e., $H=3$).}

For our evaluation, we focus on the frequency band of 3.5 GHz and geographical environment of dense urban, ordinary urban, suburban, and rural.
We use 6327 maps for training and 1590 maps for test (inference).
We consider the number of samples collected by sensors within one 3D area, i.e., $k$, to be $50$, unless otherwise specified. 
As a result, we use a very small sampling rate of $0.1\%$ for evaluation.
For each $128\times 128\times 3$ 3D area, the specific locations of $k$ radio samples are randomly generated.

We implement RadioLAM on two NVIDIA RTX 4090 GPUs.
The size (number of parameters) of RadioLAM is $0.6$ billion, unless otherwise specified. 
The training procedure is sped up by the DeepSpeed ZeRO-2~\cite{ZeRO} optimization framework.
Deep learning codes are built using PyTorch.
We set the number of DDPM inference steps as 1000, and the number of DDIM inference steps as 10.
There are 64 radio map candidates generated by the generation block for each target 2D plane in each problem instance.

To evaluate RadioLAM, we compare it with the following two interpolation-based baseline algorithms:
\begin{itemize}
    \item 3D-RBF~\cite{rbf}: It directly extends the algorithm RBF from 2D radio map construction to 3D radio map construction.
    By creating a network of basis functions around the known data points, it uses a linear combination of these basis functions to approximate the values of unknown points.

    \item 3D-kriging~\cite{kriging}: It directly extends the algorithm kriging from 2D radio map construction to 3D radio map construction. 
    It uses a semi-variogram function to describe the spatial autocorrelation and estimates the values of unknown points by weighted average of the observed data of known points.
\end{itemize}

In addition, we compare RadioLAM with the following two existing deep learning-based approaches which solve the 3D radio map construction problem:
\begin{itemize}
    \item 3D-UNet~\cite{krijestorac2021spatial}: It uses a UNet model for 3D radio map construction.
    UNet is a convolutional neural network (CNN) featuring a symmetric encoder-decoder structure. The encoder progressively compresses spatial information through downsampling, while the decoder reconstructs high-quality features via upsampling operations.

    \item 3D-DCRGAN~\cite{Hu23:TVT:3DMap}: Adapted from the GAN framework proposed in~\cite{Hu23:TVT:3DMap} for 3D radio map construction, this baseline approach removes the original algorithm's dependence on transmitters' information to maintain consistency with our problem formulation (which does not require the prior knowledge of transmitters' information).  
\end{itemize}

Note that our 3D radio map construction problem aims to estimate the radio map for a target 2D plane at height $h_t$ from radio samples in $\mathcal{S}$.
Each radio sample $s\in\mathcal{S}$ gives the RSS value located at $(x_s,y_s,h_s)$ where the height $h_s$ can be different from the target height $h_t$. 
For evaluation, we further compare RadioLAM with two existing deep learning-based methods, i.e., Autoencoder~\cite{autoencoder} and RadioUNet~\cite{RadioUNet}, both of which solve the 2D radio map construction problem from the RSS values at locations $(x_s,y_s,h_t)$ for all $s\in\mathcal{S}$.
Considering that $\{(x_s,y_s,h_s),s\in\mathcal{S}\}$ rather than $\{(x_s,y_s,h_t),s\in\mathcal{S}\}$ are inputs to our problem, the baseline methods Autoencoder and RadioUNet only represent idealized performance benchmarks rather than practical solutions for our problem.
If the practical model RadioLAM outperforms the two idealized baselines, it will be convincing that RadioLAM is effective for 3D radio map construction.

{To ensure a fair comparison, we strictly maintain identical training and test set splits for RadioLAM and all baseline methods.
More details about the simulation settings, including sets of training and test, as well as our code, are publicly available at: https://github.com/liuzhiyuan-pku/RadioLAM.}

\begin{figure*}[]
	\centering
\subfloat[Radio maps for the 2D plane at height $h_t=1.5$ m.\label{E8-1.5}]{\includegraphics[width=\linewidth]{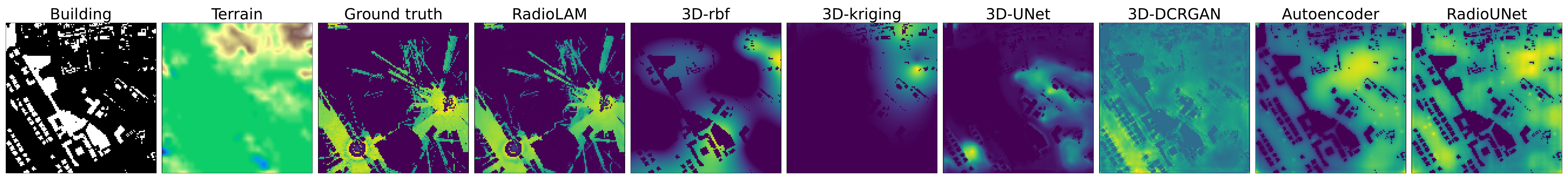}}\\
\subfloat[Radio maps for the 2D plane at height $h_t=30$ m.\label{E8-30}]{\includegraphics[width=\linewidth]{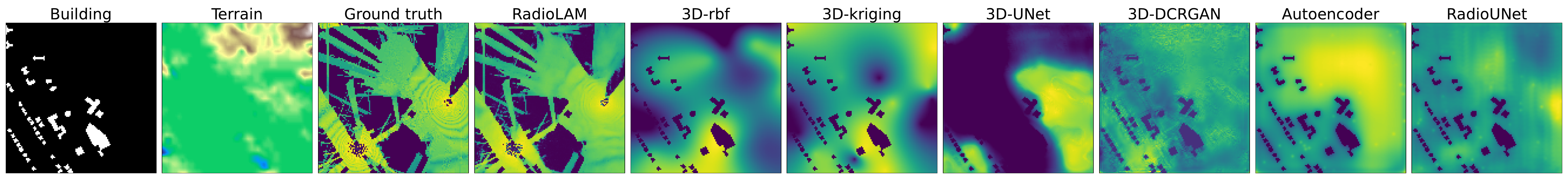}}\\
\subfloat[Radio maps for the 2D plane at height $h_t=200$ m.\label{E8-200}]{\includegraphics[width=\linewidth]{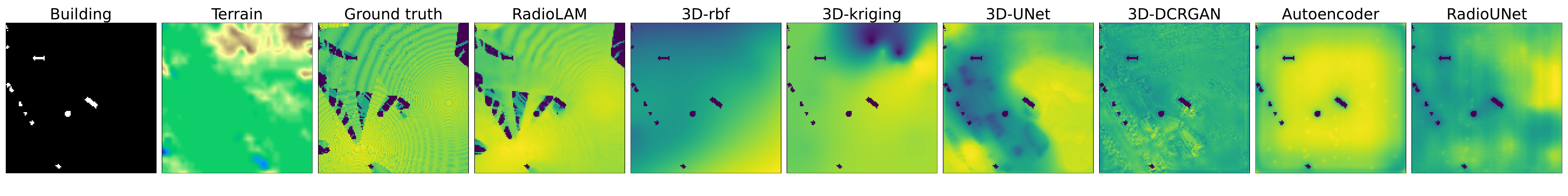}}
	\caption{Fine-grained radio maps constructed by different algorithms in the case study (ordinary urban environment).}
	\label{fig:case-study-o}
\end{figure*}

\begin{figure*}[]
	\centering
\subfloat[Radio maps for the 2D plane at height $h_t=1.5$ m.\label{E7-1.5}]{\includegraphics[width=\linewidth]{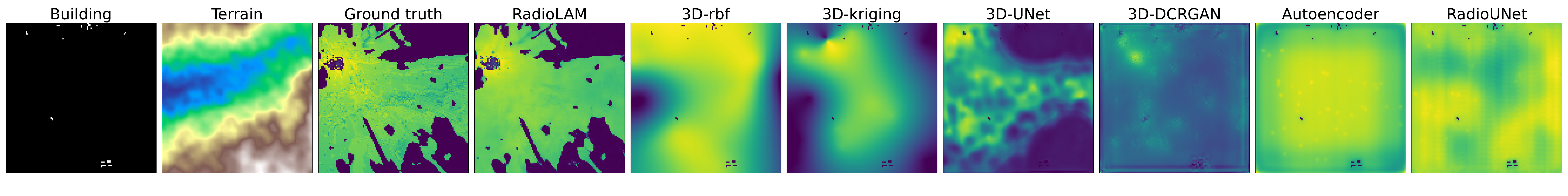}}\\
\subfloat[Radio maps for the 2D plane at height $h_t=30$ m.\label{E7-30}]{\includegraphics[width=\linewidth]{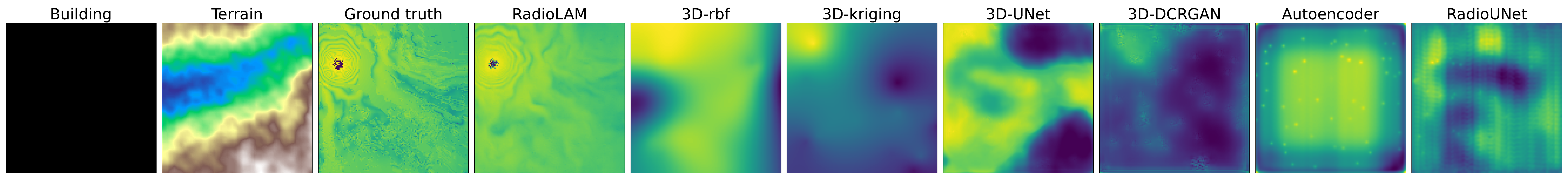}}\\
\subfloat[Radio maps for the 2D plane at height $h_t=200$ m.\label{E7-200}]{\includegraphics[width=\linewidth]{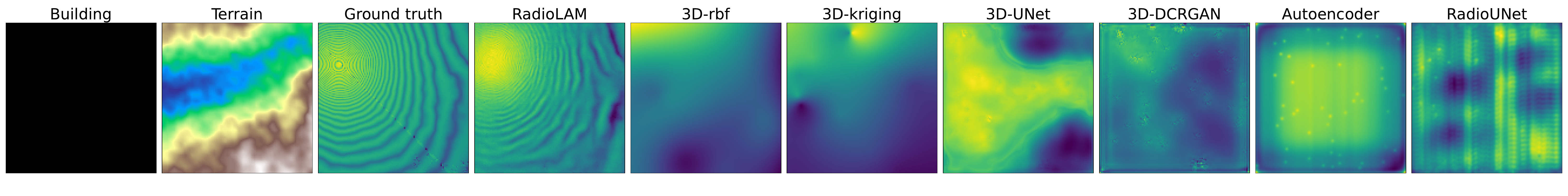}}
	\caption{Fine-grained radio maps constructed by different algorithms in the case study (rural environment).}
	\label{fig:case-study-r}
\end{figure*}

To quantify the performance of algorithms, Mean Absolute Error (MAE), MSE, and Peak Signal to Noise Ratio (PSNR) are used:
\begin{itemize}
\item MAE: It measures the average absolute difference between predicted and true values.
\begin{equation}
    \text{MAE}=\frac{1}{X\times Y}\sum^{X-1}_{x=0}\sum^{T-1}_{y=0}|I(x,y)-\hat{I}(x,y)|,
\end{equation}
where $I(x,y)$ is the true RSS value at the location $(x,y,h_t)$ and $\hat{I}(x,y)$ is the predicted RSS value at that location.

\item MSE: It is the average of the squared difference between predicted and true values.
\begin{equation}\label{eqn:mse}
    \text{MSE}=\frac{1}{X\times Y}\sum^{X-1}_{x=0}\sum^{T-1}_{y=0}[I(x,y)-\hat{I}(x,y)]^2.
\end{equation}

\item PSNR~\cite{psnr}: It estimates the ratio between the maximum RSS and the square root of MSE, and is usually expressed as a logarithmic quantity using the decibel scale.
\begin{equation}\text{PSNR}=20\cdot\log_{10}\left(\frac{\text{MAX}_I}{\sqrt{\text{MSE}}}\right),\end{equation}
where $\text{MAX}_I$ is the maximum RSS value of a point on the radio map. 
\end{itemize}

A small MAE, a small MSE, or a large PSNR implies that the constructed radio map is of high quality.

\subsection{Simulation Results}

\smallskip

\noindent {\bf Case Study\/} \ \ \
In the case study, we evaluate the performance of different algorithms across two distinct environments: an ordinary urban instance and a rural instance.
Fig.~\ref{fig:case-study-o} shows the building layouts, terrain features, and fine-grained radio maps estimated by different algorithms for an ordinary urban instance (Fig.~\ref{E8-1.5} shows the radio maps estimated at height $1.5$  m, Fig.~\ref{E8-30} shows the radio maps estimated at height $30$ m, and Fig.~\ref{E8-200} shows the radio maps estimated at height $200$ m).
Similarly, Fig.~\ref{fig:case-study-r} shows the building layouts, terrain features, and fine-grained radio maps estimated by different algorithms for a rural instance (Fig.~\ref{E7-1.5} shows the radio maps estimated at height $1.5$m, Fig.~\ref{E7-30} shows the radio maps estimated at height $30$ m, and Fig.~\ref{E7-200} shows the radio maps estimated at height $200$ m).

From the two figures, it is clear that the fine-grained radio maps constructed by RadioLAM are quite close to the ground truth (true radio maps), for both environments at all $3$ different altitudes (heights).
In sharp contrast, the radio maps constructed by any of the 6 baseline algorithms significantly deviate from the true radio maps and hence are extremely inaccurate.

\begin{table*}[]
    \centering
    \caption{Simulation results for the ordinary urban environment in the case study.}\label{tab:case}
    \begin{tabular}{c|ccc|ccc|ccc}\hline\hline 
    Performance metric & \multicolumn{3}{c|}{MAE} & \multicolumn{3}{c|}{MSE} & \multicolumn{3}{c}{PSNR} \\ \hline
        Target height $h_t$ & 1.5 m & 30 m & 200 m & 1.5 m & 30 m & 200 m & 1.5 m & 30 m & 200 m \\ \hline
        3D-RBF&0.1301 & 0.1790 & 0.1093 & 0.0398 & 0.0508 & 0.0232 & 13.998 & 12.938 & 16.340\\
3D-kriging&0.1289 & 0.1757 & 0.0686 & 0.0433 & 0.0476 & 0.0138 & 13.625 & 13.215 & 18.579\\\hline 
Autoencoder&0.1348 & 0.1399 & 0.0871 & 0.0287 & 0.0288 & 0.0111 & 15.413 & 15.396 & 19.513\\
RadioUNet&0.1381 & 0.1562 & 0.1019 & 0.0265 & 0.0288 & 0.0145 & 15.755 & 15.393 & 18.365\\\hline 
3D-UNet&0.1014 & 0.1572 & 0.0820 & 0.0286 & 0.0465 & 0.0133 & 15.423 & 13.317 & 18.741\\
3D-DCRGAN&0.1054 & 0.2169 & 0.2964 & 0.0236 & 0.0652 & 0.0960 & 16.257 & 11.856 & 10.172\\\hline 
RadioLAM&\textbf{0.0169} & \textbf{0.0252} & \textbf{0.0182} & \textbf{0.0025} & \textbf{0.0032} & \textbf{0.0013} & \textbf{25.937} & \textbf{24.849} & \textbf{28.796}
\\\hline 
\hline 
\end{tabular}
\end{table*}

Table~\ref{tab:case} presents the MAE, MSE, and PSNR results achieved by different algorithms for the ordinary urban environment.
The results indicate that in general, RadioLAM achieves over $70\%$ reduction in MAE, over $85\%$ reduction in MSE, and over $45\%$ improvement in PSNR compared to all baseline algorithms.
Interestingly, while these quantitative metrics, e.g., MAE, show that RadioLAM outperforms baseline algorithms by $70\%$, visual analysis in Fig.~\ref{fig:case-study-o} reveals a more substantial qualitative difference.
In Fig.~\ref{fig:case-study-o}, the radio map constructed by RadioLAM is very close to the true map, but those constructed by baseline algorithms significantly deviate from the true map.
This suggests potential limitations in using conventional performance metrics MAE, MSE, and PSNR for quality evaluation of radio maps.
We leave it as an important future direction to design better performance metrics for quantifying the quality of estimated radio maps.

\smallskip

\noindent {\bf More Simulations\/} \ \ \
Then we perform more simulations to evaluate RadioLAM.
Table~\ref{tab:maintable} presents the simulation results comparing RadioLAM with $6$ baseline algorithms for radio map construction of $4$ different kinds of 3D environments at $3$ different target heights.
Given one specific kind of 3D environment, we simulate 100 instances and for each instance, we use RadioLAM as well as baseline approaches to construct radio maps for 2D planes at the target heights $h_t=1.5$ m, $h_t=30$ m, and $h_t=200$ m, respectively.
The reported metrics (MAE, MSE, and PSNR) in Table~\ref{tab:maintable} represent the average values obtained from 100 instances.
{Table~\ref{tab:reduction} presents the MSE reduction results comparing the MSE value achieved by RadioLAM with that achieved by each baseline approach.
From Table~\ref{tab:maintable} and Table~\ref{tab:reduction}, we observe that RadioLAM significantly outperforms all baseline algorithms.}

\begin{table*}[]
    \centering
    \caption{Simulation results for different kinds of environments where 100 instances are simulated for each environment.}\label{tab:maintable}
    \begin{tabular}{c|ccc|ccc|ccc|ccc}
    \hline\hline\multicolumn{13}{c}{\textbf{MAE}}\\ \hline 
    3D environment & \multicolumn{3}{c|}{Suburban } & \multicolumn{3}{c|}{ Dense urban } & \multicolumn{3}{c|}{ Rural } & \multicolumn{3}{c}{ Ordinary urban } \\ \hline
        Target height $h_t$ & 1.5 m & 30 m & 200 m & 1.5 m & 30 m & 200 m & 1.5 m & 30 m & 200 m& 1.5 m & 30 m & 200 m \\ \hline
        3D-RBF&0.1782&0.1155&0.0364&0.1252&0.1501&0.0538&0.1906&0.1319&0.0416&0.1533&0.1442&0.0602\\
3D-kriging&0.1706&0.0773&0.0255&0.1055&0.1283&0.0407&0.1788&0.0837&0.0296&0.1432&0.1203&0.0453\\\hline 
Autoencoder&0.1771&0.1564&0.1405&0.1286&0.1451&0.1192&0.1904&0.1716&0.1632&0.1684&0.1628&0.1359\\
RadioUNet&0.1854&0.1315&0.1067&0.1443&0.1865&0.1629&0.1951&0.1240&0.0936&0.1743&0.1619&0.1089\\\hline 
3D-UNet&0.1916&0.1686&0.0477&0.0925&0.2609&0.0956&0.1797&0.1726&0.0494&0.1352&0.2097&0.0855\\
3D-DCRGAN&0.2438&0.1982&0.0521&0.1109&0.1760&0.0572&0.2036&0.1509&0.0421&0.1621&0.1739&0.0687\\\hline 
RadioLAM&\textbf{0.0792}&\textbf{0.0497}&\textbf{0.0165}&\textbf{0.0512}&\textbf{0.0807}&\textbf{0.0249}&\textbf{0.0848}&\textbf{0.0541}&\textbf{0.0190}&\textbf{0.0555}&\textbf{0.0684}&\textbf{0.0311}\\\hline \hline\multicolumn{13}{c}{\textbf{MSE}}\\ \hline 
    3D environment & \multicolumn{3}{c|}{Suburban } & \multicolumn{3}{c|}{ Dense urban } & \multicolumn{3}{c|}{ Rural } & \multicolumn{3}{c}{ Ordinary urban } \\ \hline
        Target height $h_t$ & 1.5 m & 30 m & 200 m & 1.5 m & 30 m & 200 m & 1.5 m & 30 m & 200 m& 1.5 m & 30 m & 200 m \\ \hline
        3D-RBF&0.0557&0.0297&0.0039&0.0379&0.0403&0.0088&0.0619&0.0351&0.0053&0.0497&0.0418&0.0126\\
3D-kriging&0.0514&0.0188&0.0023&0.0321&0.0346&0.0063&0.0546&0.0208&0.0035&0.0449&0.0340&0.0086\\\hline 
Autoencoder&0.0382&0.0281&0.0221&0.0257&0.0295&0.0172&0.0409&0.0322&0.0286&0.0363&0.0325&0.0222\\
RadioUNet&0.0427&0.0238&0.0155&0.0258&0.0416&0.0308&0.0514&0.0226&0.0115&0.0397&0.0334&0.0163\\\hline 
3D-UNet&0.0574&0.0484&0.0051&0.0244&0.0953&0.0192&0.0470&0.0505&0.0055&0.0380&0.0701&0.0137\\
3D-DCRGAN&0.0857&0.0514&0.0045&0.0324&0.0404&0.0071&0.0594&0.0345&0.0038&0.0497&0.0423&0.0096\\\hline 
RadioLAM&\textbf{0.0224}&\textbf{0.0108}&\textbf{0.0013}&\textbf{0.0153}&\textbf{0.0193}&\textbf{0.0030}&\textbf{0.0249}&\textbf{0.0123}&\textbf{0.0019}&\textbf{0.0165}&\textbf{0.0182}&\textbf{0.0056}\\\hline 
\hline\multicolumn{13}{c}{\textbf{PSNR}}\\ \hline 
    3D environment & \multicolumn{3}{c|}{Suburban } & \multicolumn{3}{c|}{ Dense urban } & \multicolumn{3}{c|}{ Rural } & \multicolumn{3}{c}{ Ordinary urban } \\ \hline
        Target height $h_t$ & 1.5 m & 30 m & 200 m & 1.5 m & 30 m & 200 m & 1.5 m & 30 m & 200 m& 1.5 m & 30 m & 200 m \\ \hline
        3D-RBF&12.541&15.267&24.119&14.208&13.948&20.559&12.081&14.546&22.741&13.038&13.787&19.000\\
3D-kriging&12.894&17.259&26.476&14.940&14.615&22.016&12.628&16.826&24.560&13.482&14.683&20.637\\\hline 
Autoencoder&14.178&15.513&16.555&15.903&15.306&17.647&13.887&14.922&15.439&14.400&14.882&16.539\\
RadioUNet&13.695&16.227&18.103&15.882&13.807&15.108&12.890&16.456&19.408&14.015&14.758&17.875\\\hline 
3D-UNet&12.411&13.153&22.933&16.124&10.209&17.173&13.275&12.964&22.600&14.207&11.543&18.647\\
3D-DCRGAN&10.672&12.893&23.484&14.899&13.934&21.510&12.262&14.621&24.254&13.038&13.738&20.193\\\hline 
RadioLAM&\textbf{16.504}&\textbf{19.675}&\textbf{28.994}&\textbf{18.149}&\textbf{17.152}&\textbf{25.237}&\textbf{16.030}&\textbf{19.101}&\textbf{27.243}&\textbf{17.830}&\textbf{17.410}&\textbf{22.546}\\\hline 
\hline\end{tabular}
\end{table*}

\begin{table*}[]
    \centering
    \caption{MSE reduction results comparing RadioLAM with baseline algorithms for different kinds of environments.}\label{tab:reduction}
    \begin{tabular}{c|ccc|ccc|ccc|ccc} \hline\hline\multicolumn{13}{c}{\textbf{MSE reduction comparing the MSE of RadioLAM with that of baseline algorithms}}\\ \hline 
    3D environment & \multicolumn{3}{c|}{Suburban } & \multicolumn{3}{c|}{ Dense urban } & \multicolumn{3}{c|}{ Rural } & \multicolumn{3}{c}{ Ordinary urban } \\ \hline
        Target height $h_t$ & 1.5 m & 30 m & 200 m & 1.5 m & 30 m & 200 m & 1.5 m & 30 m & 200 m& 1.5 m & 30 m & 200 m \\ \hline
        3D-RBF&$59.8\%$&$63.8\%$&$67.5\%$&$59.6\%$&$52.2\%$&$65.9\%$&$59.7\%$&$65.0\%$&$64.5\%$&$66.8\%$&$56.6\%$&$55.8\%$\\
3D-kriging&$56.4\%$&$42.7\%$&$44.0\%$&$52.2\%$&$44.2\%$&$52.4\%$&$54.3\%$&$40.8\%$&$46.1\%$&$63.3\%$&$46.6\%$&$35.6\%$\\\hline 
Autoencoder+&$41.5\%$&$61.6\%$&$94.3\%$&$40.4\%$&$34.6\%$&$82.6\%$&$38.9\%$&$61.8\%$&$93.4\%$&$54.6\%$&$44.1\%$&$74.9\%$\\
RadioUNet+&$47.6\%$&$54.8\%$&$91.9\%$&$40.7\%$&$53.7\%$&$90.3\%$&$51.5\%$&$45.6\%$&$83.5\%$&$58.5\%$&$45.7\%$&$65.9\%$\\\hline 
3D-UNet&$61.0\%$&$77.7\%$&$75.2\%$&$37.3\%$&$79.8\%$&$84.4\%$&$47.0\%$&$75.7\%$&$65.7\%$&$56.6\%$&$74.1\%$&$59.3\%$\\
3D-DCRGAN&$73.9\%$&$79.0\%$&$71.9\%$&$52.7\%$&$52.3\%$&$57.6\%$&$58.0\%$&$64.4\%$&$49.8\%$&$66.8\%$&$57.1\%$&$41.8\%$\\\hline
\hline\end{tabular}
\end{table*}

\begin{figure*}[]
	\centering
    \subfloat[MSE results of radio maps for 2D planes at height $h_t=1.5$ m.\label{sp-1.5}]{
		\includegraphics[width=\linewidth]{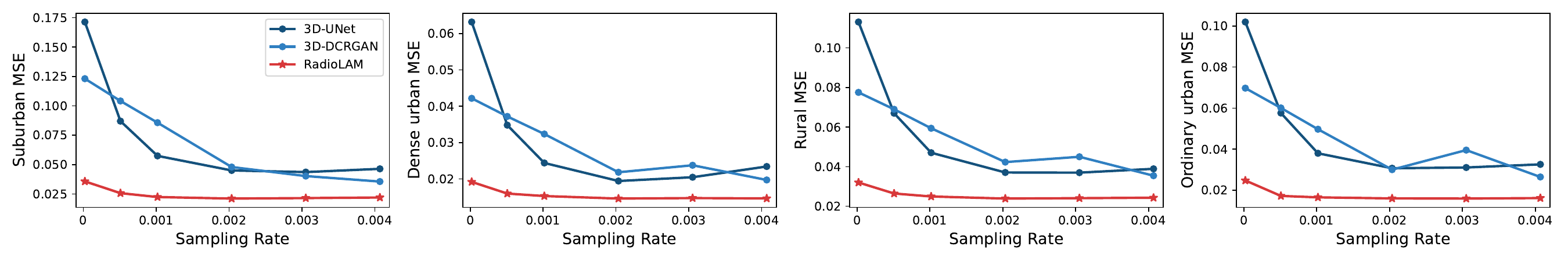}}\\
    \subfloat[MSE results of radio maps for 2D planes at height $h_t=30$ m.\label{sp-30}]{
		\includegraphics[width=\linewidth]{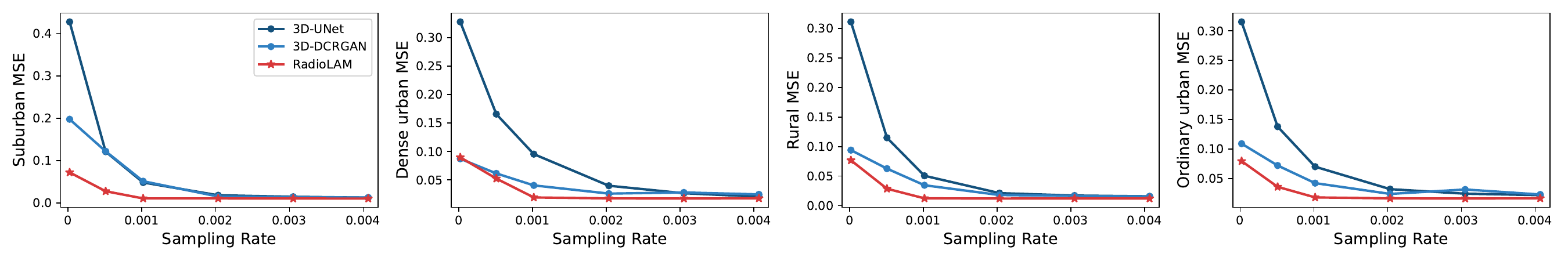}}\\
    \subfloat[MSE results of radio maps for 2D planes at height $h_t=200$ m.\label{sp-200}]{
		\includegraphics[width=\linewidth]{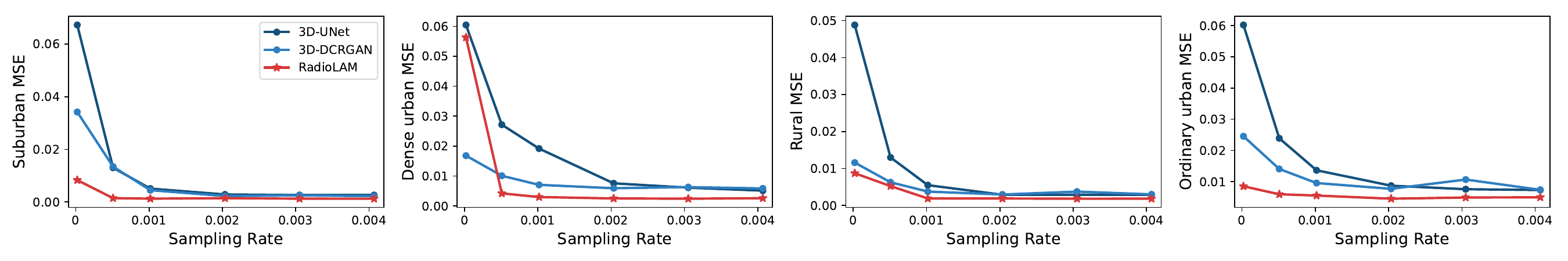}}\\
	\caption{Simulation results under varying sampling rates.}
    \label{fig:sample-rate}
\end{figure*}

Fig. \ref{fig:sample-rate} presents the  MSE results of RadioLAM, 3D-UNet, and 3D-DCRGAN, under varying sampling rates (i.e., $k/(X\cdot Y\cdot H)$).
From Fig.~\ref{fig:sample-rate}a (for $h_t=1.5$ m), we observe that the MSE achieved by RadioLAM is close to $0$ when the sampling rate is only $0.1\%$.
However, the MSE achieved by 3D-UNet and 3D-DCRGAN remains much greater than $0$ even when the sampling rate is $0.4\%$.
Similar observation can be observed from Fig.~\ref{fig:sample-rate}c for $h_t=200$ m.
From Fig.~\ref{fig:sample-rate}b (for $h_t=30$ m), the MSE achieved by RadioLAM is close to $0$ when the sampling rate is $0.1\%$, while that achieved by 3D-UNet and 3D-DCRGAN becomes close to $0$ when the sampling rate is $0.2\%$.

\smallskip
\noindent {\bf Ablation Study\/} \ \ \
RadioLAM consists of three key blocks: an augmentation block, a generation block, and an election block.
Now we use ablation studies to evaluate each of them.

\begin{table}[]
            \centering
            \caption{MSE results of simulations that evaluate the augmentation block and election block of RadioLAM.}\label{tab:ab-table}
            \begin{tabular}{cc|ccc}\hline\hline\multicolumn{5}{c}{\textbf{Suburban}}\\ \hline 
             \multicolumn{2}{c|}{Setting } & \multicolumn{3}{c}{Target height $h_t$}  \\ \hline
                 Augmentation block & Election block & 1.5 m & 30 m & 200 m \\ \hline
                 $\times$ & $\times$&0.0322&0.0200&0.0086\\
$\checkmark$ & $\times$&0.0301&0.0140&0.0019\\
$\times$ & $\checkmark$&0.0305&0.0191&0.0079\\
$\checkmark$ & $\checkmark$&\textbf{0.0224}&\textbf{0.0108}&\textbf{0.0013}\\
\hline\hline\multicolumn{5}{c}{\textbf{Dense urban}}\\ \hline 
             \multicolumn{2}{c|}{Setting } & \multicolumn{3}{c}{Target height $h_t$}  \\ \hline
                 Augmentation block & Election block & 1.5 m & 30 m & 200 m \\ \hline
                 $\times$ & $\times$&0.0229&0.0374&0.0153\\
$\checkmark$ & $\times$&0.0207&0.0291&0.0058\\
$\times$ & $\checkmark$&0.0211&0.0366&0.0153\\
$\checkmark$ & $\checkmark$&\textbf{0.0153}&\textbf{0.0193}&\textbf{0.0030}\\
\hline\hline\multicolumn{5}{c}{\textbf{Rural}}\\ \hline 
             \multicolumn{2}{c|}{Setting } & \multicolumn{3}{c}{Target height $h_t$}  \\ \hline
                 Augmentation block & Election block & 1.5 m & 30 m & 200 m \\ \hline
                 $\times$ & $\times$&0.0324&0.0202&0.0077\\
$\checkmark$ & $\times$&0.0305&0.0155&0.0023\\
$\times$ & $\checkmark$&0.0318&0.0173&0.0073\\
$\checkmark$ & $\checkmark$&\textbf{0.0249}&\textbf{0.0123}&\textbf{0.0019}\\
\hline\hline\multicolumn{5}{c}{\textbf{Ordinary urban}}\\ \hline 
             \multicolumn{2}{c|}{Setting } & \multicolumn{3}{c}{Target height $h_t$}  \\ \hline
                 Augmentation block & Election block & 1.5 m & 30 m & 200 m \\ \hline
                 $\times$ & $\times$&0.0245&0.0316&0.0131\\
$\checkmark$ & $\times$&0.0221&0.0288&0.0075\\
$\times$ & $\checkmark$&0.0234&0.0313&0.0114\\
$\checkmark$ & $\checkmark$&\textbf{0.0165}&\textbf{0.0182}&\textbf{0.0056}\\
\hline\hline\end{tabular}
         \end{table}

\begin{table}[]
            \centering
            \caption{MSE results of simulations that evaluate the free-space propagation model and Hata propagation model used by the augmentation block of RadioLAM.}\label{tab:phyab}
            \begin{tabular}{cc|ccc}\hline\hline\multicolumn{5}{c}{\textbf{Suburban}}\\ \hline 
             \multicolumn{2}{c|}{Setting } & \multicolumn{3}{c}{Target height }  \\ \hline
                 Free-space & Hata & 1.5m & 30m & 200m \\ \hline
                 $\times$ & $\times$&0.0305&0.0191&0.0079\\
$\checkmark$ & $\times$&0.0298&0.0137&0.0016\\
$\times$ & $\checkmark$&0.0278&0.0132&0.0016\\
$\checkmark$ & $\checkmark$&\textbf{0.0224}&\textbf{0.0108}&\textbf{0.0013}\\
\hline\hline\multicolumn{5}{c}{\textbf{Dense urban}}\\ \hline 
             \multicolumn{2}{c|}{Setting } & \multicolumn{3}{c}{Target height }  \\ \hline
                 Free-space & Hata & 1.5m & 30m & 200m \\ \hline
                 $\times$ & $\times$&0.0211&0.0366&0.0153\\
$\checkmark$ & $\times$&0.0211&0.0287&0.0056\\
$\times$ & $\checkmark$&0.0202&0.0291&0.0062\\
$\checkmark$ & $\checkmark$&\textbf{0.0153}&\textbf{0.0193}&\textbf{0.0030}\\
\hline\hline\multicolumn{5}{c}{\textbf{Rural}}\\ \hline 
             \multicolumn{2}{c|}{Setting } & \multicolumn{3}{c}{Target height }  \\ \hline
                 Free-space & Hata & 1.5m & 30m & 200m \\ \hline
                 $\times$ & $\times$&0.0318&0.0173&0.0073\\
$\checkmark$ & $\times$&0.0326&0.0151&0.0024\\
$\times$ & $\checkmark$&0.0315&0.0151&0.0024\\
$\checkmark$ & $\checkmark$&\textbf{0.0249}&\textbf{0.0123}&\textbf{0.0019}\\
\hline\hline\multicolumn{5}{c}{\textbf{Ordinary urban}}\\ \hline 
             \multicolumn{2}{c|}{Setting } & \multicolumn{3}{c}{Target height }  \\ \hline
                 Free-space & Hata & 1.5m & 30m & 200m \\ \hline
                 $\times$ & $\times$&0.0234&0.0313&0.0114\\
$\checkmark$ & $\times$&0.0232&0.0277&0.0078\\
$\times$ & $\checkmark$&0.0226&0.0269&0.0081\\
$\checkmark$ & $\checkmark$&\textbf{0.0165}&\textbf{0.0182}&\textbf{0.0056}\\
\hline\hline\end{tabular}
\end{table}

\begin{table}[]
            \centering
            \caption{MSE results of simulations that evaluate the generation block of RadioLAM.}\label{tab:generation}
            \begin{tabular}{c|cc|ccc}\hline\hline
            \multicolumn{6}{c}{\textbf{Suburban}}\\ \hline 
             Setting of the &\multicolumn{2}{c|}{Parameters} & \multicolumn{3}{c}{Target height $h_t$}  \\ 
                  generation block & Total & Activated & 1.5 m & 30 m & 200 m \\ \hline
                 Monolithic & $246$M & $246$M &0.0270&0.0139&0.0017\\
                 MoE-based & $609$M & $244$M &\textbf{0.0224}&\textbf{0.0108}&\textbf{0.0013}\\
                 \hline\hline
            \multicolumn{6}{c}{\textbf{Dense urban}}\\ \hline 
             Setting of the &\multicolumn{2}{c|}{Parameters} & \multicolumn{3}{c}{Target height $h_t$}  \\ 
                 generation block & Total & Activated & 1.5 m & 30 m & 200 m \\ \hline
                 Monolithic& $246$M & $246$M &0.0207&0.0292&0.0056\\
                 MoE-based& $609$M & $244$M &\textbf{0.0153}&\textbf{0.0193}&\textbf{0.0030}\\
                 \hline\hline
            \multicolumn{6}{c}{\textbf{Rural}}\\ \hline 
             Setting of the &\multicolumn{2}{c|}{Parameters} & \multicolumn{3}{c}{Target height $h_t$}  \\ 
                 generation block & Total & Activated & 1.5 m & 30 m & 200 m \\ \hline
                 Monolithic& $246$M & $246$M &0.0318&0.0155&0.0023\\
                 MoE-based& $609$M & $244$M &\textbf{0.0249}&\textbf{0.0123}&\textbf{0.0019}\\
                 \hline\hline
            \multicolumn{6}{c}{\textbf{Ordinary urban}}\\ \hline 
             Setting of the &\multicolumn{2}{c|}{Parameters} & \multicolumn{3}{c}{Target height $h_t$}  \\ 
                 generation block & Total & Activated & 1.5 m & 30 m & 200 m \\ \hline
                 Monolithic& $246$M & $246$M &0.0226&0.0277&0.0086\\
                 MoE-based& $609$M & $244$M &\textbf{0.0165}&\textbf{0.0182}&\textbf{0.0056}\\
                 \hline\hline\end{tabular}
\end{table}

Table \ref{tab:ab-table} presents an ablation study that evaluates the contribution of the augmentation block and the election block to RadioLAM. 
Given one specific kind of 3D environment, we simulate 100 instances.
This table demonstrates that:
(1) removing either the augmentation block or the election block leads to significant performance degradation, confirming both blocks' critical roles in RadioLAM;
(2) the augmentation block exhibits greater impact on the output quality of RadioLAM than the election block, as evidenced by more substantial performance drops when this block is removed.

\begin{figure*}[]
	\centering
\includegraphics[width=\linewidth]{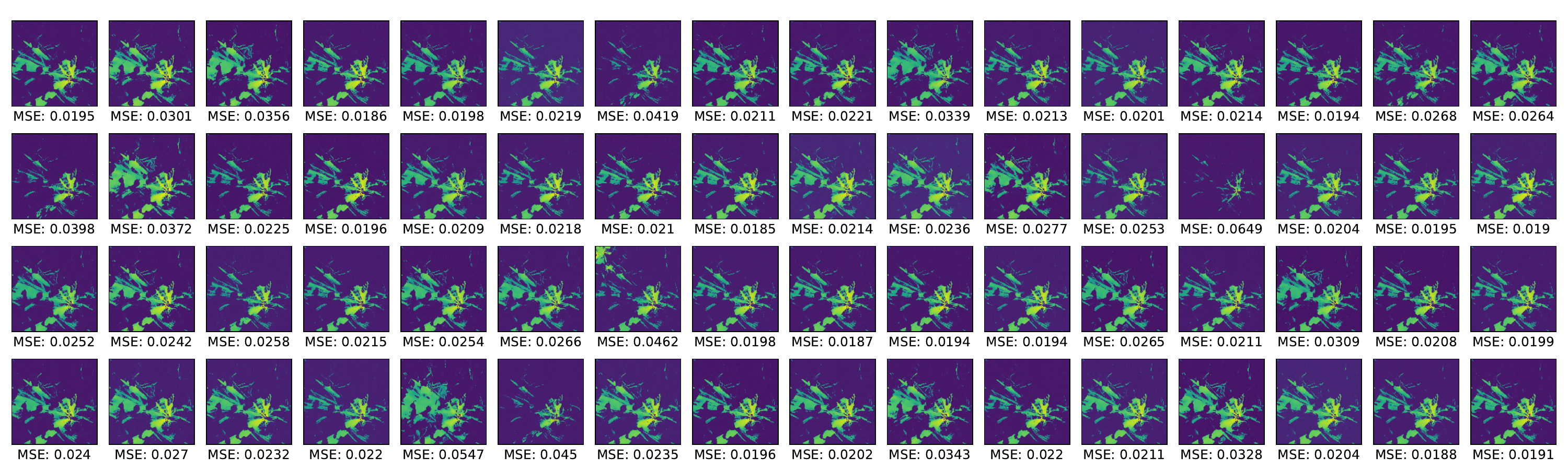}
    \caption{The set of 64 candidate radio maps (as well as the achieved MSE values) generated by the generation block of RadioLAM in one simulation instance when the target height is $h_t=1.5$ m.}
    \label{fig:div}
\end{figure*}

The augmentation block projects the set of multi-altitude radio samples collected by sensors onto the target 2D plane using Equation~\eqref{eqn:aug}, which integrates the free-space and Hata radio propagation models.
Next, we use an ablation study to evaluate the contribution of the free-space model and Hata model to RadioLAM, with the results presented in Table~\ref{tab:phyab}.
Given one specific kind of 3D environment, we simulate 100 instances.
This table demonstrates that both the free-space and Hata models are essential for effective augmentation.
Radio maps constructed using only one of the models exhibit quality levels similar to those produced without any augmentation, and are significantly inferior to maps constructed by RadioLAM when both models are employed together.

{Moreover, we use simulations to show that the candidate radio maps generated by the generation block are diverse, and hence an election is needed to identify the best map.
For the instance simulated in Fig.~\ref{fig:case-study-o} at the target height $h_t=1.5$ m, now Fig. \ref{fig:div} illustrates the candidate set of 64 fine-grained radio maps generated by the generation block along with their corresponding MSE values.
Fig. \ref{fig:div} demonstrates a significant variation in map quality, with the best map achieving an MSE of $0.0185$ compared to $0.0649$ for the poorest map, indicating that it is necessary to have an election block in RadioLAM.
}

The generation block is the core of RadioLAM and has a MoE architecture.
In this design, each expert in MoE specializes in generating radio maps for a distinct propagation environment.
To validate this design, we conduct simulations to compare our MoE-based model with a monolithic baseline using a single diffusion model for all environments.
The number of parameters in the monolithic model is $246$ million, consistent with the number of activated parameters in our MoE-based model. 
Table~\ref{tab:generation} shows the simulation results, clearly demonstrating the superior performance of our proposed MoE-based generation over the monolithic generation. 

\smallskip

\begin{figure*}[]
	\centering
\includegraphics[width=0.9\linewidth]{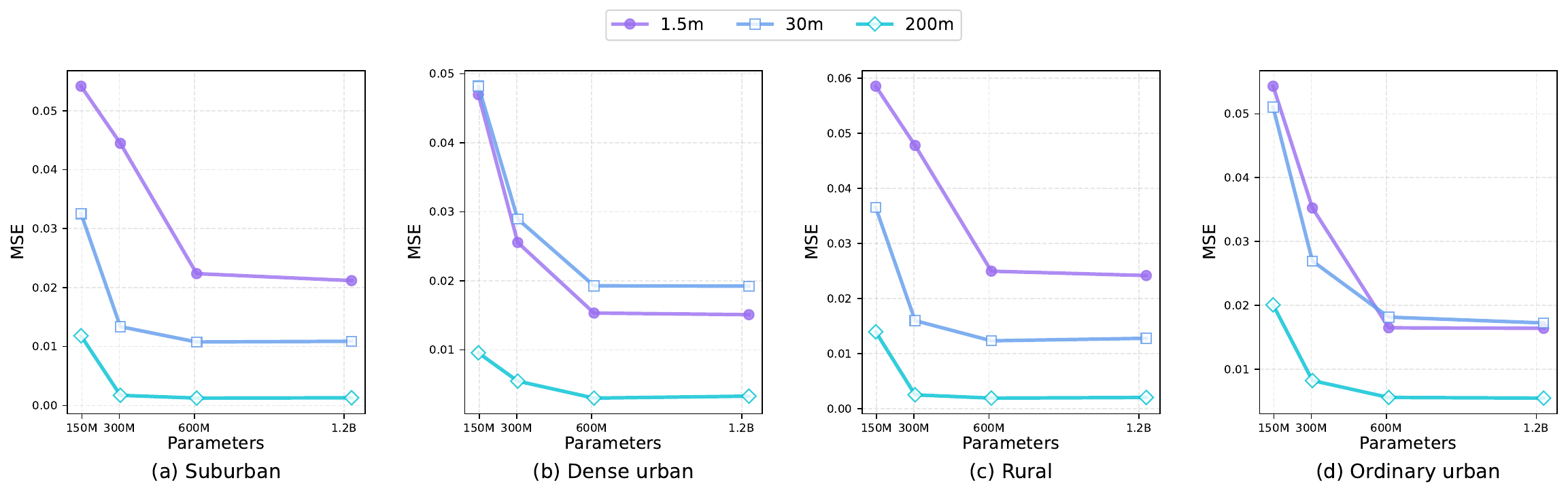}	
    \caption{Performance of RadioLAM under varying model sizes (numbers of parameters).}
    \label{fig:diffP}
\end{figure*}

\noindent {\bf Model Size Analysis\/} \ \ \
Here we evaluate the performance of RadioLAM from the perspective of its model size (number of parameters).
First, we evaluate RadioLAM under varying model sizes.
Fig.~\ref{fig:diffP} demonstrates the MSE results of RadioLAM when the number of parameters is $150$ million, $300$ million, $0.6$ billion, and $1.2$ billion, respectively.
From the figure, we observe that when the number of parameters is smaller than $0.6$ billion, the performance improvement of RadioLAM is evident as the number of parameters grows.
However, when the number of parameters is greater than $0.6$ billion, the performance improvement of RadioLAM is minor as the number of parameters increases.

{Due to its inherent complexity, RadioLAM requires much more parameters compared to the baselines Autoencoder, RadioUNet, 3D-UNet and 3D-DCRGAN. 
A critical question arises:}
{\emph{Would baseline approaches perform comparably or even outperform RadioLAM if their parameter count were increased to match or exceed that of RadioLAM?}}
{We address this question empirically through comprehensive evaluations.}

{We enhance each of the baseline models – Autoencoder, RadioUNet, 3D-UNet and 3D-DCRGAN – by increasing their depth (number of layers), expanding their width (channel dimensions), and training them with parameter counts comparable to or exceeding those of RadioLAM.
The resulting improved variants are denoted as Autoencoder+, RadioUNet+, 3D-UNet+ and 3D-DCRGAN+, respectively.
For each algorithm, the number of parameters and the MSE results of simulations are shown in Table \ref{tab:para}. 
From the table, it is clear that the radio map constructed by RadioLAM is significantly better than that constructed by any of the enhanced baselines (Autoencoder+, RadioUNet+, 3D-UNet+ and 3D-DCRGAN+), considering that the MSE achieved by RadioLAM is significantly lower than that achieved by any of the enhanced baselines.}

{Overall, even when the baseline approaches are scaled up to match or exceed RadioLAM’s model size, RadioLAM maintains superior performance.}

\begin{table}[]
            \centering
            \caption{MSE results and model sizes of different algorithms for different kinds of environments.}\label{tab:para}
            \begin{tabular}{cc|ccc}\hline\hline\multicolumn{5}{c}{\textbf{Suburban}}\\ \hline 
             \multicolumn{2}{c|}{Setting } & \multicolumn{3}{c}{Target height }  \\ \hline
                 Method & Parameters & 1.5 m & 30 m & 200 m \\ \hline
                 Autoencoder+ & $906M$&0.0463&0.0164&0.0070\\
RadioUNet+& $682M$&0.0441&0.0228&0.0149\\\hline
3D-UNet+& $671M$&0.0396&0.0151&0.0020\\
3D-DCRGAN+ & $671M$&0.0531&0.0250&0.0025\\\hline
RadioLAM & $609M$&\textbf{0.0224}&\textbf{0.0108}&\textbf{0.0013}\\
\hline\hline\multicolumn{5}{c}{\textbf{Dense urban}}\\ \hline 
             \multicolumn{2}{c|}{Setting } & \multicolumn{3}{c}{Target height }  \\ \hline
                 Method & Parameters & 1.5 m & 30 m & 200 m \\ \hline
                 Autoencoder+ & $906M$&0.0482&0.0326&0.0071\\
RadioUNet+& $682M$&0.0459&0.0298&0.0114\\\hline
3D-UNet+& $671M$&0.0212&0.0246&0.0058\\
3D-DCRGAN+ & $671M$&0.0240&0.0340&0.0068\\\hline
RadioLAM & $609M$&\textbf{0.0153}&\textbf{0.0193}&\textbf{0.0030}\\
\hline\hline\multicolumn{5}{c}{\textbf{Rural}}\\ \hline 
             \multicolumn{2}{c|}{Setting } & \multicolumn{3}{c}{Target height }  \\ \hline
                 Method & Parameters & 1.5 m & 30 m & 200 m \\ \hline
                 Autoencoder+ & $906M$&0.0501&0.0195&0.0105\\
RadioUNet+& $682M$&0.0490&0.0218&0.0144\\\hline
3D-UNet+& $671M$&0.0412&0.0194&0.0029\\
3D-DCRGAN+ & $671M$&0.0473&0.0306&0.0033\\\hline
RadioLAM & $609M$&\textbf{0.0249}&\textbf{0.0123}&\textbf{0.0019}\\
\hline\hline\multicolumn{5}{c}{\textbf{Ordinary urban}}\\ \hline 
             \multicolumn{2}{c|}{Setting } & \multicolumn{3}{c}{Target height }  \\ \hline
                 Method & Parameters & 1.5 m & 30 m & 200 m \\ \hline
                 Autoencoder+ & $906M$&0.0570&0.0338&0.0124\\
RadioUNet+& $682M$&0.0499&0.0336&0.0178\\\hline
3D-UNet+& $671M$&0.0287&0.0237&0.0083\\
3D-DCRGAN+ & $671M$&0.0345&0.0331&0.0091\\\hline
RadioLAM & $609M$&\textbf{0.0165}&\textbf{0.0182}&\textbf{0.0056}\\
\hline\hline\end{tabular}
         \end{table}

\subsection{{Real-World Experimental Settings}}

{Finally, we use the real-world open-source dataset \emph{AERPAW Dataset-22} \cite{singh2024dataset22} to evaluate RadioLAM.
In this dataset, a drone equipped with a Samsung S21 smartphone flew a semi-circular trajectory at an altitude of $35.7$ m around a private Ericsson AERPAW base station. 
The smartphone logged 4G cellular Key Performance Indicators (KPIs), including RSS, and these logs were correlated with the drone’s position based on timestamps.
For the evaluation,  we focus on a $1.28$ km $\times$ $1.28$ km area encompassing the region with the highest density of RSS measurements.
This area is discretized into a $128$ $\times$ $128$ grid, with each grid cell corresponding to a small area of $10$ m $\times$ $10$ m. 
RSS measurements within the same grid cell are averaged to produce a single value.
Following this data preprocessing, we get $781$ distinct sampling points with associated RSS values.}

{Given the limited size of \emph{AERPAW Dataset-22}, it is not feasible to use it to effectively train a large generative model such as RadioLAM. 
Therefore, in this paper, we adopt a zero-shot evaluation setting.
Specifically, all models, including RadioLAM and the baseline algorithms, are first trained on the large-scale simulated dataset \emph{SpectrumNet}.
They are then evaluated on the small-scale real-world \emph{AERPAW Dataset-22} without further fine-tuning.
This zero-shot evaluation setting is commonly employed in fields such as 
computer vision \cite{zero-shot-denoising,zero-shot-video,zero-shot-generalization}, particularly when real-world data is scarce or when assessing the generalization capability of proposed models.
During inference, $n$ samples from the $781$ available points in \emph{AERPAW Dataset-22} are used as inputs to the models.
The models' predictions for the remaining $(781-n)$ samples are compared against their ground-truth values, with the performance evaluated using MSE and MAE.}

\subsection{{Real-World Experimental Results}}

\begin{table}[]
\centering
            \caption{Real-world experimental results based on the data from the AERPAW UAV testbed, using a 5G Ericsson network.}\label{tab:real}
\resizebox{1\linewidth}{!}{%
\begin{tabular}{c|cc|cc|cc}
\hline\hline
\specialcell{Number of samples\\that are inputs to\\different algorithms} & \multicolumn{2}{c|}{25}           & \multicolumn{2}{c|}{50}           & \multicolumn{2}{c}{75}            \\ \hline
     Metrics & MSE             & MAE             & MSE             & MAE             & MSE             & MAE             \\ \hline
     Free-space & 0.0626& 0.1974 &0.0543 & 0.1863 & 0.0747	& 0.2279\\ \hline
3D-RBF      & 0.0661          & 0.1943          & 0.0550          & 0.1540          & 0.0658          & 0.1568          \\
3D-kriging  & 0.0236&	0.1206          & 0.0371	&0.1486          & 0.0275	&0.1252          \\ \hline
Autoencoder & 0.0406          & 0.1723          & 0.0408          & 0.1731          & 0.0403          & 0.1724          \\
RadioUNet   & 0.0409          & 0.1728          & 0.0338          & 0.1573          & 0.0291          & 0.1455          \\ \hline
3D-UNet     & 0.1172          & 0.3058          & 0.1167          & 0.3051          & 0.1157          & 0.3043          \\
3D-DCRGAN   & 0.0342          & 0.1593          & 0.0232          & 0.1265          & 0.0192          & 0.1143          \\ \hline
RadioLAM    & \textbf{0.0176} & \textbf{0.1092} & \textbf{0.0146} & \textbf{0.0978} & \textbf{0.0150} & \textbf{0.0995} \\ \hline\hline
\end{tabular}}
\end{table}

{In addition to the baseline algorithms evaluated previously, we also compare RadioLAM with a free-space baseline which leverages the free-space propagation model to construct radio maps.
The real-world experimental results are presented in Table~\ref{tab:real}.
Comparing Table~\ref{tab:real} with Table~\ref{tab:maintable}, we observe that the radio maps constructed by RadioLAM on the real-world \emph{AERPAW Dataset-22} are less accurate than those constructed by RadioLAM on the simulated \emph{SpectrumNet} (note that the sampling rate on \emph{AERPAW Dataset-22} is greater than that on \emph{SpectrumNet}), which is intuitive since RadioLAM was trained on \emph{SpectrumNet} rather than \emph{AERPAW Dataset-22}.
Although RadioLAM becomes less accurate on \emph{AERPAW Dataset-22}, from Table~\ref{tab:real}, it is clear that RadioLAM consistently and significantly outperforms all baseline algorithms, particularly at low sampling rates.
Specifically, RadioLAM achieves a lower MSE than any of the $7$ baseline methods across all evaluated values of $n$.
Moreover, when $n=25$, the MSE achieved by RadioLAM is nearly $50\%$ lower than that achieved by any of the $6$ baselines including free-space, 3D-RBF, Autoencoder, RadioUNet, 3D-UNet, and 3D-DCRGAN.
Besides, when $n=50$, the MSE of RadioLAM is over $60\%$ lower than that of the baseline 3D-kriging.}

\section{Conclusion}\label{Conclusion}
The problem of fine-grained 3D radio map construction aims to construct high-resolution radio maps for 2D planes at desired altitudes (target heights) within a 3D area of interest, based on radio samples collected at different locations within that 3D area.
This problem is challenging due to the ultra-sparse sampling, i.e., the number of collected samples is significantly smaller than the high resolution of radio maps to be estimated.
In this paper, we leverage the creative power and the strong generalization capability of generative LAM to address the ultra-sparse sampling challenge, and propose RadioLAM for fine-grained 3D radio map construction.
RadioLAM consists of three key blocks:
1) an augmentation block, which leverages the principle of radio propagation to project radio samples collected at varying altitudes onto the target 2D plane at the desired altitude; 
2) a generation block, which employs an LAM under an MoE architecture to generate a candidate set of diverse fine-grained radio maps for the target 2D plane, based on projections of radio samples;
3) an election block, which utilizes the radio propagation principle as a guide to find the best map from the candidate set, and uses TTA to dynamically adjust the noise level of LAM during inference.

Extensive simulations based on the open-source dataset \emph{SpectrumNet} show that for different kinds of radio propagation environments, including rural, suburban, ordinary urban, and dense urban, 
1) RadioLAM produces high-quality, fine-grained radio maps from an ultra-small sampling rate of $0.1\%$;
2) fine-grained radio maps generated by RadioLAM are over $50\%$ better than those generated by SOTA under the $0.1\%$ sampling rate;
3) RadioLAM requires at most one fourth of the sampling rate needed by SOTA to construct fine-grained radio maps with comparable qualities.
{Moreover, experiments based on the open-source real-world dataset \emph{AERPAW} also demonstrate that the radio maps achieved by RadioLAM are much more accurate than those achieved by SOTA.}
Potential future research includes the design of 3D radio map construction algorithms with strong theoretical performance guarantees and radio map temporal prediction from currently collected radio samples.

\bibliographystyle{IEEEtran}
\bibliography{RadioLAMRef}

\begin{thebibliography}{10}
\providecommand{\url}[1]{#1}
\csname url@samestyle\endcsname
\providecommand{\newblock}{\relax}
\providecommand{\bibinfo}[2]{#2}
\providecommand{\BIBentrySTDinterwordspacing}{\spaceskip=0pt\relax}
\providecommand{\BIBentryALTinterwordstretchfactor}{4}
\providecommand{\BIBentryALTinterwordspacing}{\spaceskip=\fontdimen2\font plus
\BIBentryALTinterwordstretchfactor\fontdimen3\font minus \fontdimen4\font\relax}
\providecommand{\BIBforeignlanguage}[2]{{%
\expandafter\ifx\csname l@#1\endcsname\relax
\typeout{** WARNING: IEEEtran.bst: No hyphenation pattern has been}%
\typeout{** loaded for the language `#1'. Using the pattern for}%
\typeout{** the default language instead.}%
\else
\language=\csname l@#1\endcsname
\fi
#2}}
\providecommand{\BIBdecl}{\relax}
\BIBdecl

\bibitem{RME}
D.~Romero and S.-J. Kim, ``Radio map estimation: A data-driven approach to spectrum cartography,'' \emph{IEEE Signal Processing Magazine}, vol.~39, no.~6, pp. 53--72, 2022.

\bibitem{Peng24:TCE:Access}
Y.~Peng, Y.~Li, Y.~Guo, D.~Zhang, F.~Khan, R.~Alturki, and B.~Alshawi, ``A blockchain-based distributed collaborative sensing and spectrum access approach for consumer electronics,'' \emph{IEEE Transactions on Consumer Electronics}, 2025, early access.

\bibitem{Matar24:JSAC:Sharing}
A.~S. Matar and X.~Shen, ``Joint optimization of user association, power control, and dynamic spectrum sharing for integrated aerial-terrestrial network,'' \emph{IEEE Journal on Selected Areas in Communications}, vol.~43, no.~1, pp. 396--409, 2025.

\bibitem{Huang24:TWC:inference}
J.~Huang, L.~Lian, D.~Wen, Y.~Zhou, F.~Wang, W.~Wang, and Y.~Shi, ``Dynamic {UAV}-assisted cooperative edge {AI} inference,'' \emph{IEEE Transactions on Wireless Communications}, vol.~24, no.~1, pp. 615--628, 2024.

\bibitem{Zhu24:IoTJ:surveillance}
H.~Zhu, U.~Demirbaga, G.~S. Aujla, L.~Shi, and P.~Zhang, ``Explainable edge {AI} framework for {IoD}-assisted aerial surveillance in extreme scenarios,'' \emph{IEEE Internet of Things Journal}, vol.~12, no.~5, pp. 4570--4578, 2025.

\bibitem{Qi24:TVT:SAR}
S.~Qi, B.~Lin, Y.~Deng, X.~Chen, and Y.~Fang, ``Minimizing maximum latency of task offloading for multi-{UAV}-assisted maritime search and rescue,'' \emph{IEEE Transactions on Vehicular Technology}, vol.~73, no.~9, pp. 13\,625--13\,638, 2024.

\bibitem{Ri24:NC:inspection}
S.~Ri, J.~Ye, N.~Toyama, and N.~Ogura, ``Drone-based displacement measurement of infrastructures utilizing phase information,'' \emph{Nature Communications}, vol.~15, no.~1, 2024.

\bibitem{ray-tracing}
K.~Rizk, J.-F. Wagen, and F.~Gardiol, ``Two-dimensional ray-tracing modeling for propagation prediction in microcellular environments,'' \emph{IEEE Transactions on Vehicular Technology}, vol.~46, no.~2, pp. 508--518, 1997.

\bibitem{dominant-path}
R.~Wahl, G.~Wölfle, P.~Wildbolz, and F.~Landstorfer, ``Dominant path prediction model for urban scenarios,'' in \emph{Proc. of IST Mobile and Wireless Communications, Dresden, Germany, June 19-23}, 2005.

\bibitem{eg2}
J.~A. Bazerque and G.~B. Giannakis, ``Distributed spectrum sensing for cognitive radio networks by exploiting sparsity,'' \emph{IEEE Transactions on Signal Processing}, vol.~58, no.~3, pp. 1847--1862, 2010.

\bibitem{eg}
T.~Zugno, M.~Drago, M.~Giordani, M.~Polese, and M.~Zorzi, ``Toward standardization of millimeter-wave vehicle-to-vehicle networks: Open challenges and performance evaluation,'' \emph{IEEE Communications Magazine}, vol.~58, no.~9, pp. 79--85, 2020.

\bibitem{rbf}
C.~M. Bishop, \emph{Neural Networks for Pattern Recognition}.\hskip 1em plus 0.5em minus 0.4em\relax New York, NY, USA: Oxford University Press, Inc., 1995.

\bibitem{spline}
J.~A. Bazerque, G.~Mateos, and G.~B. Giannakis, ``Group-lasso on splines for spectrum cartography,'' \emph{IEEE Transactions on Signal Processing}, vol.~59, no.~10, pp. 4648--4663, 2011.

\bibitem{kriging}
G.~Boccolini, G.~Hernández-Peñaloza, and B.~Beferull-Lozano, ``Wireless sensor network for spectrum cartography based on kriging interpolation,'' in \emph{Proc. of IEEE PIMRC, Sydney, Australia, September 09-12}, 2012.

\bibitem{RadioUNet}
R.~Levie, {\c{C}}.~Yapar, G.~Kutyniok, and G.~Caire, ``{RadioUNet}: Fast radio map estimation with convolutional neural networks,'' \emph{IEEE Transactions on Wireless Communications}, vol.~20, no.~6, pp. 4001--4015, 2021.

\bibitem{AE}
Y.~Teganya and D.~Romero, ``Deep completion autoencoders for radio map estimation,'' \emph{IEEE Transactions on Wireless Communications}, vol.~21, no.~3, pp. 1710--1724, 2022.

\bibitem{ResNet}
K.~He, X.~Zhang, S.~Ren, and J.~Sun, ``Deep residual learning for image recognition,'' in \emph{Proc. of CVPR, Las Vegas, USA, June 27-30}, 2016.

\bibitem{Liu25:ArXiv:diffusion}
Z.~Liu, S.~Zhang, Q.~Liu, H.~Zhang, and L.~Song, ``{WiFi-Diffusion}: Achieving fine-grained {WiFi} radio map estimation with ultra-low sampling rate by diffusion models,'' \emph{arXiv:2503.12004v2}, 2025.

\bibitem{FEED_ref}
M.~Ayadi, A.~Ben~Zineb, and S.~Tabbane, ``A {UHF} path loss model using learning machine for heterogeneous networks,'' \emph{IEEE Transactions on Antennas and Propagation}, vol.~65, no.~7, pp. 3675--3683, 2017.

\bibitem{GAN_ref}
A.~Creswell, T.~White, V.~Dumoulin, K.~Arulkumaran, B.~Sengupta, and A.~A. Bharath, ``Generative adversarial networks: An overview,'' \emph{IEEE Signal Processing Magazine}, vol.~35, no.~1, pp. 53--65, 2018.

\bibitem{RadioGAT}
X.~Li, S.~Zhang, H.~Li, X.~Li, L.~Xu, H.~Xu, H.~Mei, G.~Zhu, N.~Qi, and M.~Xiao, ``{RadioGAT}: A joint model-based and data-driven framework for multi-band radiomap reconstruction via graph attention networks,'' \emph{IEEE Transactions on Wireless Communications}, vol.~23, no.~11, pp. 1--1, 2024.

\bibitem{Hu23:TVT:3DMap}
T.~Hu, Y.~Huang, J.~Chen, Q.~Wu, and Z.~Gong, ``{3D} radio map reconstruction based on generative adversarial networks under constrained aircraft trajectories,'' \emph{IEEE Transactions on Vehicular Technology}, vol.~72, no.~6, pp. 8250--8255, 2023.

\bibitem{zhao20253d}
L.~Zhao, Z.~Fei, X.~Wang, J.~Luo, and Z.~Zheng, ``{3D-RadioDiff}: An altitude-conditioned diffusion model for {3D} radio map construction,'' \emph{IEEE Wireless Communications Letters}, 2025, early access.

\bibitem{krijestorac2021spatial}
E.~Krijestorac, S.~Hanna, and D.~Cabric, ``Spatial signal strength prediction using {3D} maps and deep learning,'' in \emph{Proc. of IEEE ICC}, Virtual Conference, June 14-23, 2021.

\bibitem{Chen15:Letter:3DMap}
Z.~Chen, H.~Wang, and D.~Guo, ``{3D} radio map estimation based on active measurement trajectory selection,'' \emph{IEEE Wireless Communications Letters}, 2025, early access.

\bibitem{Zhang24:TCCN:3DMap}
S.~Zhang, S.~Jiang, W.~Lin, Z.~Fang, K.~Liu, H.~Zhang, and K.~Chen, ``Generative {AI} on {SpectrumNet}: An open benchmark of multiband {3D} radio maps,'' \emph{IEEE Transactions on Cognitive Communications and Networking}, vol.~11, no.~2, pp. 886--901, 2025.

\bibitem{IEEE80222}
``{IEEE} standard - information technology-telecommunications and information exchange between systems-wireless regional area networks-specific requirements-part 22: Cognitive wireless ran mac and phy specifications: Policies and procedures for operation in the bands that allow spectrum sharing where the communications devices may opportunistically operate in the spectrum of primary service,'' \emph{{IEEE} Std 802.22-2019 (Revision of {IEEE} Std 802.22-2011)}, pp. 1--1465, 2020.

\bibitem{EttusB200mini}
\BIBentryALTinterwordspacing
{Ettus Research, a National Instruments Company}, \emph{{USRP B200mini Series Data Sheet}}, Ettus Research, Santa Clara, CA, USA, 2022, revision: v1.0. [Online]. Available: \url{https://kb.ettus.com/images/6/64/USRP\_B200mini\_Data\_Sheet.pdf}
\BIBentrySTDinterwordspacing

\bibitem{DATID-3D}
G.~Kim and S.~Y. Chun, ``{DATID-3D}: Diversity-preserved domain adaptation using text-to-image diffusion for {3D} generative model,'' in \emph{Proc. of CVPR, Vancouver, Canada, June 17-24}, 2023, pp. 14\,203--14\,213.

\bibitem{topodiff}
Y.~Zhang, Y.~Liu, Z.~Ma, M.~Li, C.~Xu, and H.~Gong, ``Improving diffusion-based protein backbone generation with global-geometry-aware latent encoding,'' \emph{Nature Machine Intelligence}, vol.~7, no.~7, pp. 1104--1118, 7 2025.

\bibitem{Self-Consistency}
X.~Wang, J.~Wei, D.~Schuurmans, Q.~V. Le, E.~H. Chi, S.~Narang, A.~Chowdhery, and D.~Zhou, ``Self-consistency improves chain of thought reasoning in language models,'' in \emph{Proc. of ICLR, Kigali, Rwanda, May 1-5}, 2023.

\bibitem{Friis}
H.~Friis, ``A note on a simple transmission formula,'' \emph{Proceedings of the IRE}, vol.~34, no.~5, pp. 254--256, 1946.

\bibitem{Hata}
P.~E. Mogensen, J.~Wigard \emph{et~al.}, ``{COST} action 231: Digital mobile radio towards future generation systems, final report,'' European Commission, Luxembourg, COST Action 231 EUR 18957, 1999.

\bibitem{LM1963}
D.~W. Marquardt, ``An algorithm for least-squares estimation of nonlinear parameters,'' \emph{Journal of the Society for Industrial and Applied Mathematics}, vol.~11, no.~2, pp. 431--441, 1963.

\bibitem{DeepSeekMoE}
\BIBentryALTinterwordspacing
D.~Dai, C.~Deng, C.~Zhao, R.~X. Xu, H.~Gao, D.~Chen, J.~Li, W.~Zeng, X.~Yu, Y.~Wu, Z.~Xie, Y.~K. Li, P.~Huang, F.~Luo, C.~Ruan, Z.~Sui, and W.~Liang, ``{DeepSeekMoE}: Towards ultimate expert specialization in mixture-of-experts language models,'' 2024. [Online]. Available: \url{https://arxiv.org/abs/2401.06066}
\BIBentrySTDinterwordspacing

\bibitem{DDPM}
J.~Ho, A.~Jain, and P.~Abbeel, ``Denoising diffusion probabilistic models,'' in \emph{Proc. of NeurIPS, virtual only conference, December 06-12}, 2020.

\bibitem{CFG}
J.~Ho and T.~Salimans, ``Classifier-free diffusion guidance,'' in \emph{Proc. of NeurIPS, virtual only conference, December 06-14}, 2021.

\bibitem{DDIM}
J.~Song, C.~Meng, and S.~Ermon, ``Denoising diffusion implicit models,'' in \emph{Proc. of ICLR, virtual only conference, May 03-07}, 2021.

\bibitem{TimeDiff}
L.~Shen and J.~Kwok, ``Non-autoregressive conditional diffusion models for time series prediction,'' in \emph{Proc. of ICML, Honolulu, Hawaii, July 23-29}, 2023.

\bibitem{ARMD}
J.~Gao, Q.~Cao, and Y.~Chen, ``Auto-regressive moving diffusion models for time series forecasting,'' in \emph{Proc. of AAAI, Philadelphia, Pennsylvania, February 25 - March 29}, 2025.

\bibitem{MADM}
T.~Chen, J.~Hou, Y.~Zhou, H.~Xie, X.~Chen, Q.~Liu, X.~Guo, M.~Xia, J.~S. Duncan, C.~Liu, and B.~Zhou, ``{2.5D} multi-view averaging diffusion model for {3D} medical image translation: Application to low-count {PET} reconstruction with {CT}-less attenuation correction,'' \emph{IEEE Transactions on Medical Imaging}, vol.~44, no.~11, pp. 4239--4250, 2025.

\bibitem{bestMSE}
B.~Fesl, B.~B{\"o}ck, F.~Strasser, M.~Baur, M.~Joham, and W.~Utschick, ``On the asymptotic mean square error optimality of diffusion models,'' in \emph{Proc. of AISTATS, Mai Khao, Thailand, May 03-05}, 2025.

\bibitem{openstreetmap}
{OpenStreetMap Contributor Terms}, ``{OpenStreetMap},'' \url{https://www.openstreetmap.org/}, 2024.

\bibitem{ZeRO}
S.~Rajbhandari, J.~Rasley, O.~Ruwase, and Y.~He, ``{ZeRO}: memory optimizations toward training trillion parameter models,'' in \emph{Proceedings of the International Conference for High Performance Computing, Networking, Storage and Analysis}, ser. SC '20.\hskip 1em plus 0.5em minus 0.4em\relax IEEE Press, 2020.

\bibitem{autoencoder}
W.~Locke, N.~Lokhmachev, Y.~Huang, and X.~Li, ``Radio map estimation with deep dual path autoencoders and skip connection learning,'' in \emph{Proc. of IEEE PIMRC, Toronto, Canada, September 05-08}, 2023.

\bibitem{psnr}
A.~Horé and D.~Ziou, ``Image quality metrics: {PSNR} vs. {SSIM},'' in \emph{Proc. of IEEE ICPR, Istanbul, Turkey, August 23-26}, 2010.

\bibitem{singh2024dataset22}
\BIBentryALTinterwordspacing
S.~Singh, ``Dataset-22: Android-based {4G LTE} measurements for semi-circular {UAV} trajectory around a private {AERPAW} base station,'' AERPAW, 2024, dataset contains LTE KPIs (RSRP, RSRQ, RSSI) and UAV position data collected using Samsung S21 with PawPrints Android App. Data available in CSV format on Google Drive, with post-processing scripts on GitHub. [Online]. Available: \url{https://aerpaw.org/dataset/dataset-22-android-based-4g-lte-measurements-for-semi-circular-uav-trajectory-around-a-private-aerpaw-base-station/}
\BIBentrySTDinterwordspacing

\bibitem{zero-shot-denoising}
Y.~Wang, J.~Yu, and J.~Zhang, ``Zero-shot image restoration using denoising diffusion null-space model,'' in \emph{Proc. of ICLR, Kigali, Rwanda, May 01-05}, 2023.

\bibitem{zero-shot-video}
L.~Khachatryan, A.~Movsisyan, V.~Tadevosyan, R.~Henschel, Z.~Wang, S.~Navasardyan, and H.~Shi, ``Text2video-zero: Text-to-image diffusion models are zero-shot video generators,'' in \emph{Proc. of ICCV, Paris, France, October 02-06}, 2023.

\bibitem{zero-shot-generalization}
P.~Li, Z.~Li, H.~Zhang, and J.~Bian, ``On the generalization properties of diffusion models,'' in \emph{Proc. of NeurIPS, New Orleans, Louisiana, December 10-16}, 2023.

\end{thebibliography}

\section*{Biographies}
\begin{IEEEbiography}[{\includegraphics[width=1in,height=1.25in,clip,keepaspectratio]{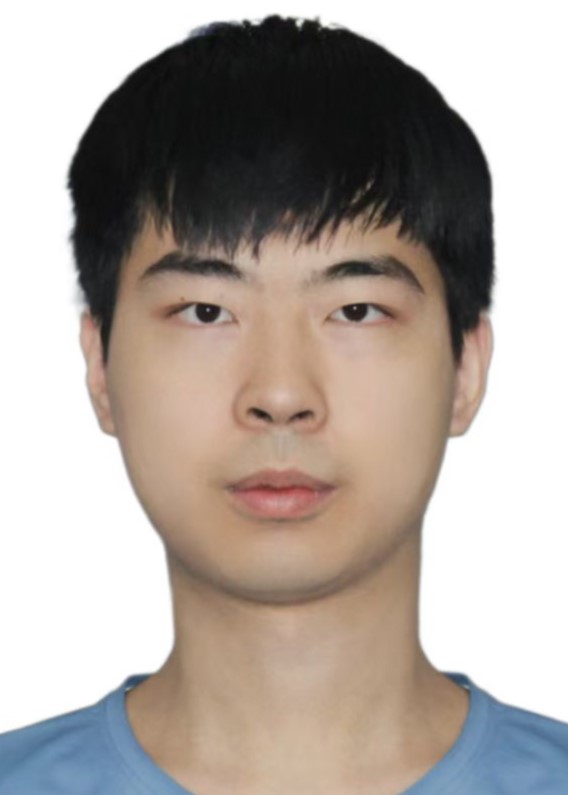}}] {Zhiyuan Liu} (Student Member, IEEE) received the B.S. degree in the School of Artificial Intelligence, Beijing University of Posts and Telecommunications, Beijing, China, in 2024. He is currently pursuing the M.S. degree with the School of Electronic and Computer Engineering, Peking University Shenzhen Graduate School. His current research interests include AI for communications and generative AI.
\end{IEEEbiography}

\begin{IEEEbiography}
[{\includegraphics[width=1in,height=1.25in,clip,keepaspectratio]{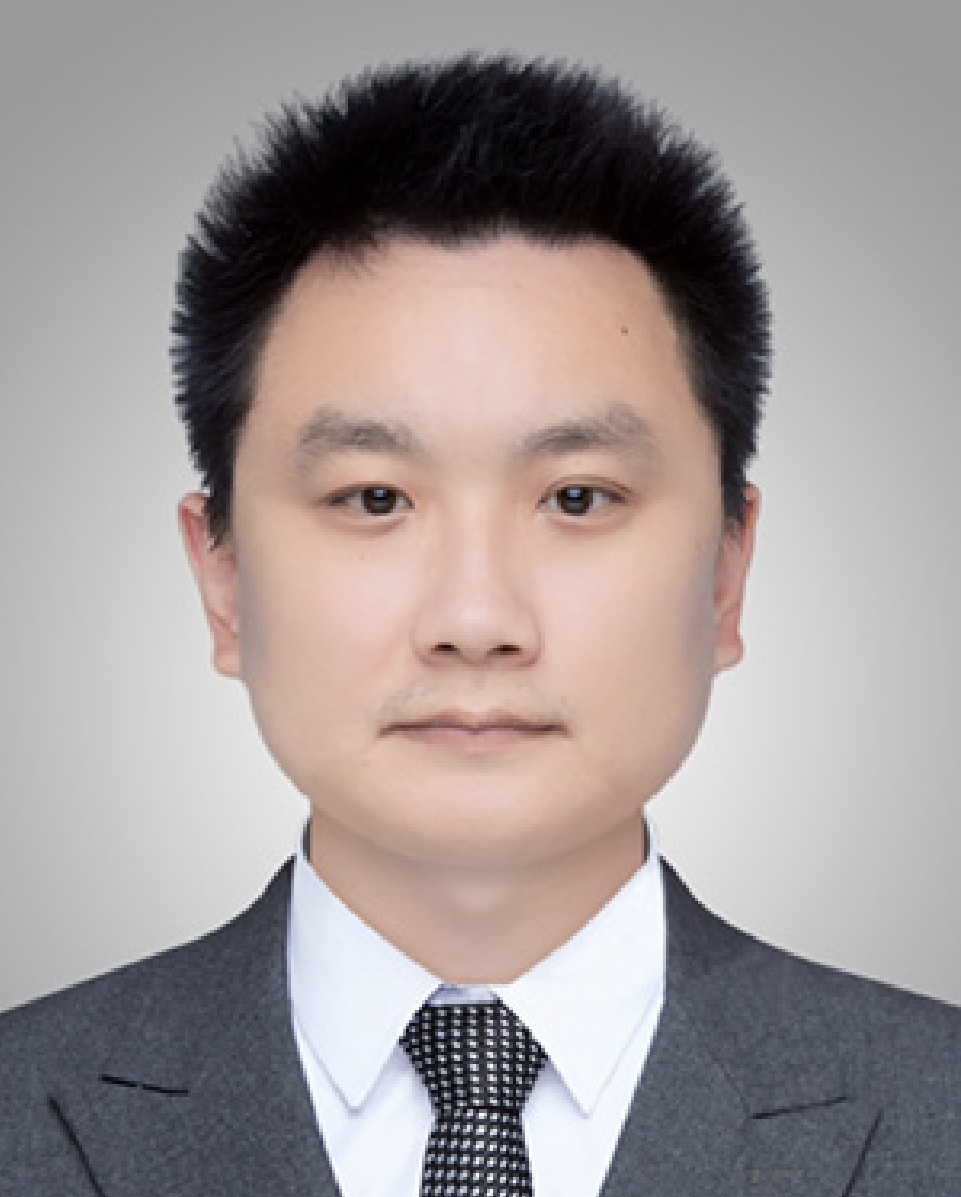}}]
{Qingyu Liu}(Member, IEEE) is currently an Assistant Professor with the School of Electronic and Computer Engineering, Peking University Shenzhen Graduate School, where he joined in May 2023. He received the Ph.D. degree in computer engineering from Virginia Tech, Blacksburg, VA, USA, in 2019. Prior to joining Peking University, he was a Postdoc and then a Research Assistant Professor of the Department of Electrical and Computer Engineering with Virginia Tech, from September 2019 to May 2023. His research interests include generative AI for wireless communication, AI RAN, low latency wireless communication, and intelligent transportation. He has been serving on TPC of IEEE INFOCOM since 2021, and was awarded as a Distinguished Member of the INFOCOM TPC in 2023. He is currently an Associate Editor for IEEE Internet of Things Journal. He was the Secretary for the IEEE Communication Society (ComSoc) Asia Pacific Board (APB) from 2024 to 2025, and now serves as a Co-Chair for the IEEE ComSoc APB Chapters Coordination Committee.
\end{IEEEbiography}

\begin{IEEEbiography}[{\includegraphics[width=1in,height=1.25in,clip,keepaspectratio]{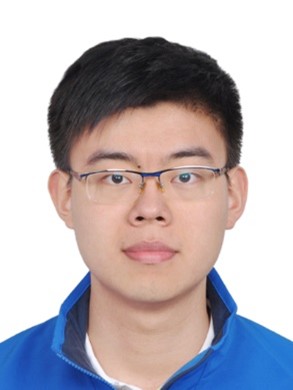}}] {Shuhang Zhang} (Member, IEEE)  received the B.S.
and Ph.D. degrees in electronic engineering from the School of Electrical Engineering and Computer Science, Peking University, Beijing, China, in 2016 and 2021, respectively. 
Since 2025, he has been with the School of Electronics, Peking University, as an Assistant Professor. 
Before joining Peking University, he was with the Peng Cheng Laboratory, as an Assistant Research Professor, from 2023 to 2025, and Huawei Technology Company Ltd., from 2021 to 2023. 
He has published over 30 papers in IEEE/ACM journals, including multiple ESI hot papers and ESI highly cited papers. 
His current research interests include space-air-ground integrated networks and artificial intelligence. 
He has won the 2021 IEEE ComSoc Heinrich Hertz Award, the 2021 IEEE ComSoc Asia Pacific Outstanding Paper Award, and the 2019 First Prize of IEEE ComSoc Student Competition. 
He is an Associate Editor of
IEEE Internet of Things Journal.
\end{IEEEbiography}

\begin{IEEEbiography}[{\includegraphics[width=1in,height=1.25in,clip,keepaspectratio]{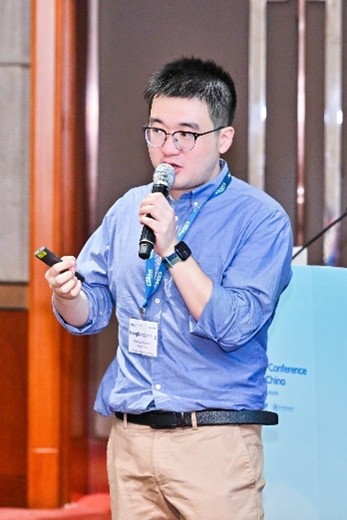}}] {Hongliang Zhang} (Member, IEEE)  is currently an Endowed Boya Young Fellow Assistant Professor with the School of Electronics, Peking University.
His current research interests include intelligent surfaces, aerial access networks, and the Internet of Things. 
He was a recipient of the IEEE ComSoc Asia Pacific Outstanding Young Researcher Award, the IEEE ComSoc Best Tutorial Paper Award, the IEEE Comsoc Heinrich Hertz Award for Best Communications Letters, the IEEE Neal Shepherd Memorial Best Propagation Paper Award, the IEEE ComSoc Asia Pacific Outstanding Paper Award, the IEEE GLOBECOM Best Paper Award, and the IEEE/CIC ICCC Best Demo Award. 
He is currently an Editor of IEEE Transactions on Cognitive Communications and Networking, IEEE Internet of Things Journal, IEEE Transactions on Vehicular Technology, IEEE Communications Letters, and IET Communications. He is an Exemplary Editor of IEEE
Communications Letters in 2023.
\end{IEEEbiography}

\begin{IEEEbiography}[{\includegraphics[width=1in,height=1.25in,clip,keepaspectratio]{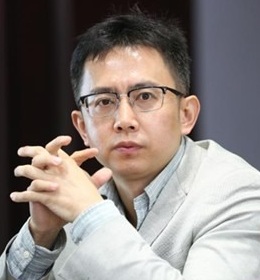}}] {Lingyang Song}(Fellow, IEEE)  received the Ph.D. degree from the University of York, U.K., in 2007.
He was a Research Fellow with the University of Oslo, Norway, until rejoining Philips Research, U.K., in March 2008. 
In May 2009, he joined the School of Electronics Engineering and Computer Science, Peking University, where he is currently a Boya Distinguished Professor. 
His main research interests include wireless communications, mobile computing, and machine learning. 
He is the coauthor of many awards, including the IEEE Leonard G. Abraham Prize in 2016, IEEE ICC 2015, IEEE ICC 2014, and IEEE Globecom 2014; and the Best Demo Award in ACM MobiHoc 2015. 
He received the National Science Fund for Distinguished Young Scholars in 2017 and the First Prize in Natural Science Award of Ministry of Education of China in 2017. 
He is a Clarivate Analytics Highly Cited Researcher. 
He has served as an IEEE ComSoc Distinguished Lecturer (2015–2018), an Area Editor for IEEE Transactions on Vehicular Technology (since 2019), and the Director of IEEE ComSoc Asia Pacific Board (2024–2025).
\end{IEEEbiography}

\end{document}